\documentclass[aps,prd,twocolumn,superscriptaddress,floatfix,showpacs]{revtex4}
\usepackage{graphics}
\usepackage{graphicx}
\usepackage{epsfig}
\usepackage{amsmath}
\usepackage{amsfonts}
\usepackage{bm}
\usepackage{url}
\usepackage{float}
\usepackage{multirow}

\def\beq{\begin{equation}}
\def\eeq{\end{equation}}
\def\bea{\begin{eqnarray}}
\def\eea{\end{eqnarray}}

\def\gsim{ \lower .75ex \hbox{$\sim$} \llap{\raise .27ex \hbox{$>$}} }
\def\lsim{ \lower .75ex\hbox{$\sim$} \llap{\raise .27ex \hbox{$<$}} }

\def\m{\mu}
\def\n{\nu}
\def\lm{\lambda^{\m}}
\def\lnu{\lambda^{\nu}}
\DeclareMathOperator{\sinc}{sinc}

\begin{document}
%% \normalsize  
\input epsf.tex

\title{Detecting compact galactic binaries using a hybrid swarm-based algorithm}
\author{ Yann \surname{Bouffanais}}
\email[]{bouffana@apc.in2p3.fr}
\author{Edward K. \surname{Porter}}
\email[]{porter@apc.in2p3.fr}
\vspace{1cm}
\affiliation{Fran\c{c}ois Arago Center, APC, UMR 7164, Universit\'e Paris 7 Denis Diderot,\\ 10, rue Alice Domon et L\'{e}onie Duquet, 75205 Paris Cedex 13, France}
\vspace{1cm}
\begin{abstract}
Compact binaries in our galaxy are expected to be one of the main sources of gravitational waves for the future eLISA mission. During the mission lifetime, many thousands
of galactic binaries should be individually resolved. However, the identification of the sources, and the extraction of the signal parameters in a noisy environment are real 
challenges for data analysis. So far, stochastic searches have proven to be the most successful for this problem. In this work we present the first application of a swarm-based algorithm combining Particle Swarm Optimization and Differential Evolution. These algorithms have been shown to converge faster to global solutions on complicated likelihood surfaces
than other stochastic methods.  We first demonstrate the effectiveness of the algorithm for the case of a 
single binary in a 1 mHz search bandwidth.  This interesting problem gave the algorithm plenty of opportunity to fail, as it can be easier to find a strong noise peak rather than the
signal itself.  After a successful detection of a fictitious low-frequency source, as well as the verification binary RXJ0806.3+1527, we then applied the algorithm to the detection of multiple binaries, over different search bandwidths, in the cases of low and mild source confusion.  In all cases, we
show that we can successfully identify the sources, and recover the true parameters within a 99\% credible interval.
\end{abstract}

\maketitle
%%%%%%%%%%%%%%%%%%%%%%%%%%%%%%%%%%%%%%%%%%%%%%%%%%%%%%%%%%%%%%%%%%%%%%%%%%%%%%%%
%%%%%%%%%%%%%%%%%%%%%%%%%%%%%
%%%%%%%%%%%%%%%%%%%%%%%%%%%%%%%%%%% Introduction 
%%%%%%%%%%%%%%%%%%%%%%%%%%%%%%%%%%%%%%%%%%%%%%%%%%%%%%%%%%%%
%%%%%%%%%%%%%%%%%%%%%%%%%%%%%%%%%%%%%%%%%%%%%%%%%%%%%%%%%%%%%%%%%%%%%%%%%%%%%%%%
%%%%%%%%%%%%%%%%%%%%%%%%%%%%%

\section{Introduction}
Compact binary systems, composed of white dwarves (WDs), neutron stars (NSs) and stellar mass black holes (BHs), are thought to be a major source of gravitational waves (GWs) in our
galaxy.  Estimations from population synthesis models predict a possible number of close to 70 million galactic binaries in the data at any one time~\cite{nelemans_2001,ruiter_2010}.  The binaries of interest for space-based GW astronomy
are composed of two compact stars with orbitals periods shorter than one hour. These binaries can be sorted into two categories: detached and interacting. In the case of detached 
binaries, there is no mass transfer between the two objects, and their dynamics is essentially governed by GW emission. Most galactic binary (GB) systems will be composed of binary WDs,   but there will also be binary NS and binary BH systems.  We should also expect a number of mixed binaries composed of different permutations compact object type.  
Interacting binaries are characterized by a transfer of momentum via mass accretion. GWs 
emitted by these objects would give crucial information on their complex dynamics. There are two main categories of interacting binaries: AM CVn systems where the accretor and the companion are 
WDs, and ultra compact x-ray binaries where the accretor is a NS, while the donor is a WD~\cite{Nelemans:2013yg}. Due to their weak interaction with matter, the 
observation of these objects with GWs should give us new insights into their nature and would add complementary information to the current electromagnetic (EM) observations. In fact, 
the GWs emitted by these sources carry important information on the distances, masses, and nature of the binary system (i.e. detached or interacting with a change in frequency). All of these 
observations can then be used in order to better understand the formation and distribution of compact objects in our Galaxy~\cite{nelemans_astrophysics_2009, kilic_ultra-compact_2013}.

However, identifying the parameters of all of these simultaneous sources is a real challenge in terms of data analysis. The first difficulty is the weak interaction between GWs and the detector, leading to very small wave amplitudes. As a consequence, all the signals will be hidden in the noise of the instrument, and only a Fourier analysis reveals the features of the signal in the noise background. Another issue is closely connected to the nature of gravitational waves and is often called the \textit{Cocktail party problem} 
~\cite{cornish_lisa_2005}. Unlike EM observations, a GW detector measures the signals in all directions simultaneously. The advantage of this 
simultaneous detection is the ability to constantly observe and gather information on the sources. The drawback is the confusion between signals: two binaries with extremely close frequencies cannot be analyzed separately if the difference between their frequencies is smaller than the smallest accessible frequency bin $\Delta f = 1/T_{obs}$, where 
$T_{obs}$ is the length of the observation period. Thus, we know that in certain frequency bins, there will be more than one binary, and it will be impossible to separate the two signals: only 
the one with the highest SNR will be recovered. While source confusion is a problem for the eLISA mission, we do not expect to have the same confusion noise from a GB foreground as with  
the LISA configuration due to the reduced size of the observatory~\cite{amaro-seoane_elisa:_2012}. Furthermore, our ability to individually resolve sources comes from the fact
that the power in a single binary is spread over several frequency bins due to the Doppler modulation coming from the orbit of the detector around the Sun~\cite{cornish_lisa_2003}. This modulation frequency is denoted as $f_{m} = 1/yr$.

Regarding these challenges, it has been necessary to develop algorithms dedicated specifically to parameter identification for GWs. There have been two major types of algorithms tested in the framework of space-based GW data analysis: grid based and stochastic searches. Grid-based algorithms use a uniform distribution of templates over the parameter space, designed to achieve a given degree of minimal match with the sources. They can either be in the form of a hierarchical search, where regions of interest are 
identified separately with refinement of the grid \cite{Blaut:2009si}, or in the form of an N-source search where metatemplates of multiple binaries are computed \cite{whelan_searching_2010}. However, given the high computational cost of computing a uniform grid of templates in a high-dimension parameter space 
\cite{cornish_detecting_2005}, stochastic approaches have been favored in the field. A variety of such algorithms have been tested, including genetic algorithms 
\cite{crowder_lisa_2006}, tomographic reconstruction \cite{mohanty_tomographic_2005} and Markov chain Monte Carlo based algorithms \cite{trias_studying_2009,crowder_solution_2006,cornish_porter_2007_2,
cornish_porter_2007_3,vecchio_2004,vecchio_2006,veitch_2008,
cornish_tests_2007,littenberg_2009}. Among these algorithms, the MCMC-based pipelines have been the most 
successful so far, especially the Block Annealed Metropolis Hastings that proved to be the most suited for detecting a full population of binaries in the Galaxy \cite{crowder_solution_2006}. 

Though much progress has been made in the GB source search, it is still necessary to look for even faster and more efficient algorithms in order to 
be ready to analyse real data when the mission is launched.  In this article we present a first work on a swarm-based search algorithm that combines Particle Swarm Optimization (PSO)\cite{kennedy_1995,shi_1998} and Differential Evolution (DE) \cite{storn_1997} during the search phase, as well as DE and Markov Chain Monte Carlo (MCMC) \cite{metropolis_1953,hastings_1970} routines for the parameter estimation phase.  While MCMC-based algorithms have performed well, their convergence is very much
dependent on the choice of an efficient proposal distribution and annealing strategies.  Once we are close to the global solution, the convergence of these methods is quite
fast.  However,  at a distant point in parameter space, these algorithms
may find themselves stuck in a random walk with slow convergence.  Both PSO and DE have been shown to have faster convergence properties than MCMC methods~\cite{Omran_2008,prasad_cosmological_2012},
especially when the likelihood surface is known to have a high population of secondary maxima.

eLISA is a future ESA L3 project dedicated to the detection of GWs using space-based interferometry \cite{amaro-seoane_elisa:_2012}. So far there has been no direct detection of GWs, but only indirect evidence from observations of binary pulsars \cite{hulse_1975}. The eLISA mission is 
being constructed as a complementary observatory to current Earth-based projects such as Advanced LIGO and Advanced VIRGO~\cite{TheLIGOScientific:2014jea,TheVirgo:2014hva}, as well as pulsar timing arrays~\cite{pta}. The eLISA frequency bandwidth of
 detection is set between $10^{-5}$ and $1$ Hz. In this bandwidth, we should detect a number of sources such as GBs, supermassive black hole binaries, extreme
  mass ratio inspirals, cosmic (super)strings, and possibly, a stochastic cosmological background. Among these sources, the vast majority will be GBs. In its current configuration, eLISA is a three-body constellation defining a two-arm interferometer of length $L = 10^{6}$ km. The constellation will orbit the Sun at a distance of 1 AU and with an angle of $20^{\circ}$ with respect to the Earth. Due to 
  the cartwheel movement of the spacecrafts, various modulations play a part in the phase of the signals measured by the detector. In terms of GB sources, their 
  frequencies span the band between $10^{-4}$ and $10^{-2}$ Hz, and it is believed that the number of resolvable galactic binaries will be close to 4000 in the first year, with an additional $10^3$ per year of observation~\cite{amaro-seoane_elisa:_2012,nissanke_gravitational-wave_2012}. 

This article is structured as follows. In Sec.~\ref{section_one}, we review the important concepts of Bayesian inference and their application to the detection of gravitational waves for ultra-compact galactic binaries. In Sec.~\ref{section_two}, we present the general features of the methods that have been used in our search algorithm.
 In Sec.~\ref{section_three}, we present the design of our search algorithm and how we benchmarked its performance on a single source search. In Sec.~\ref{section_four}, we present the performances of our search algorithm for the analysis of two multiple sources data sets.

\section{Data analysis methodology for gravitational wave astronomy}
\label{section_one}
The main difficulty in GW astronomy comes from the fact that the output of the detector is a superposition of millions of GW signals, plus a number of noise 
contributions.   A standard technique for extracting signals from noisy data is matched filtering. The idea behind this method is as follows : given a theoretical
waveform model, or template, parametrized by astrophysically motivated parameters $\{\lm\}$, what is the parameter set that maximizes the correlation 
between the template and a possible individual GW signal in the data? 

Once a signal is detected, our next goal is to estimate the parameters for the best fit template.  A number of recent studies have demonstrated that the 
Fisher information matrix is an unreliable tool in GW astronomy due to the fact that in a number of cases, the posterior distributions in the parameter
errors are non-Gaussian (a prerequisite for using the Fisher matrix in the first place)~\cite{porter_fisher_2015}.  With this in mind, our goal is to carry out a Bayesian analysis
when conducting the parameter estimation study. 

\subsection{A Bayesian framework for data analysis}
\label{Intro_DataAnalysis}
If we consider a time-series signal $s(t)$ coming from the output of an experiment, we assume that $s(t)$ contains both a noise-like part $n(t)$, related to noise coming from experimental or physical backgrounds, and a signal of interest $h(t;\lm)$. This signal depends on a parameter set $\{\lm\}$ that describes the physics of the phenomenon.  Thus,
we can write
\begin{equation}
s(t) = h(t; \lambda^{\mu}) + n(t).
\end{equation}
Bayesian inference is quickly becoming a useful tool in GW astronomy for testing model hypothesis, i.e. given a data set $s(t)$, a template $h(t;\lm)$, and a noise
model $S_n(f)$, what is the posterior probability distribution $p(\lm|s)$?  One can construct the posterior probability distribution via Bayes' theorem:
\beq
p(\lm |s) = \frac{\pi(\lm)p(s|\lm)}{p(s)}.
\eeq
The prior probability $\pi(\lm)$ reflects the \emph{a priori} knowledge of the experiment before evaluating the data,  $p(s|\lm) = {\mathcal L}(\lm)$ is the likelihood function, 
 and $p(s)$ is the marginalized likelihood or ``evidence" given by
\beq
p(s) = \int\, d\lm\, \pi(\lm)p(s|\lm).
\eeq
We define the likelihood, ${\mathcal L}(\lm)$, given the output, $s(t)$, and a GW template, $h(t)$, as
\begin{equation}
\mathcal{L}(\lm) = \exp\left[ - \frac{1}{2} \left< s -h(\lm) | s - h(\lm)\right>\right],
\label{likelihood_def}
\end{equation}
where the angular brackets define a noise-weighted scalar product defined on the manifold of all possible signals. More explicitly, given two time domain signals $h(t;\lambda_{1}^{\m})$ and $s(t;\lambda_{2}^{\m})$, we can write the scalar product as
\begin{equation}
\left<h\left|s\right.\right> =2\int_{0}^{\infty}\frac{df}{S_{n}(f)}\,\left[ \tilde{h}(f)\tilde{s}^{*}(f) +  \tilde{h}^{*}(f)\tilde{s}(f) \right],
\label{eqn:scalarprod}
\end{equation}
where $\tilde{h}(f)$ is the Fourier transform of the time domain signal $h(t)$, an asterisk denotes a complex conjugate and $S_{n} (f)$ is the one-sided noise power spectral density of the noise.  In this study, we use the eLISA observatory as our detector of choice.  In its current configuration, the one-sided noise power 
spectral density is given by~\cite{amaro-seoane_elisa:_2012}
\bea
S_n(f) & = & \frac{1}{4L^2} \left[ S_n^{fxd} + 2 S_n^{\text{pos}} \left( 2+\cos^2\left(\frac{f}{f_*}\right)\right)  \right. \nonumber \\
 & &  + 8 S_n^{\text{acc}} \left( 1+ \cos^2\left(\frac{f}{f_*}\right)\right)  \nonumber \\
 & & \left. \times \left( \frac{1}{(2\pi f)^4} + \frac{(2\pi 10^{-4})^2}{(2\pi f)^6}\right) \right],
\eea
where $L=10^9$m is the arm length of the observatory, $S_n^{fxd} = 6.28\times10^{-23}\,\text{m}^2/\text{Hz}$ is a frequency independent
 fixed level noise in the detector, $S_n^{\text{pos}} (f)= 1.21\times10^{-22}\, \text{m}^2/\text{Hz}$ is the position noise and
$S_n^{\text{acc}} (f)= 9\times10^{-30}\,\text{m}^2/(\text{s}^4\text{Hz})$ is the acceleration noise.  We highlight the fact that our expression 
for the noise also contains a red-noise component at frequencies of $f\leq10^{-4}$ Hz. 
    
Complementary to the likelihood function, we can also define the signal-to-noise ratio (SNR) as  
\begin{equation}
\rho = \dfrac{\langle s \mid h \rangle}{\sqrt{\langle h \mid h\rangle}}.
\end{equation}

\subsection{Time domain response for galactic binaries}
\label{Galactic_Waveform}
The strain of the GW as seen by the detector can be written, in the low frequency approximation (LFA)~\cite{cutler_1998}, as a linear combination of the two GW wave polarizations $h_{+,\times}(t)$ and the detector beam pattern functions $F^{+,\times}(t)$,
\begin{equation}
h(t) = h_{+}(t) F^{+}(t) + h_{\times}(t) F^{\times}(t).
\label{WaweformEq}
\end{equation}
The LFA is valid when the GW wavelength is greater than the size of the detector, or conversely, where the frequency of the wave is inferior to the mean transfer frequency $f^{*} = c/(2 \pi L)$\cite{cornish_rubbo2003,krolac_2004}. For eLISA, this corresponds to a frequency of $f^{*}\approx 10^{-2}$ Hz.

In the case of circular monochromatic GBs, and in the framework of general relativity, the two polarizations of the GW in Eq.~\eqref{WaweformEq} are given by
\begin{eqnarray}
h_{+} (t) &=& A (1 + \cos^{2} (\iota) ) \cos (\Phi (t) + \varphi_{0}), \\
h_{\times} (t) &=& -2A \cos (\iota) \sin (\Phi (t) + \varphi_{0}),
\end{eqnarray}
where $A$ is the amplitude of the wave, $\iota$ is the inclination of the orbital plane, $\varphi_{0}$ is the initial phase and $\Phi (t)$ is the phase of the wave. For a monochromatic binary, the phase can be expressed as
\begin{equation}
\Phi (t) = 2 \pi f_{0} \left[ t + R_{\oplus} \sin (\theta) \cos (2 \pi f_{m}t - \phi) \right],	
\label{Phase}
\end{equation}
where $f_{0}$ is the GW frequency (twice the value of the orbital frequency), $\theta$ is the colatitude, $\phi$ is the longitude and $R_{\oplus}=1AU$ is the radius of eLISA orbit. The phase differs from a simple monochromatic process due to the motion of the detector with respect to the source, inducing a Doppler motion contribution to the phase.
The beam pattern functions of the detector, $F^{+,\times}(t)$, are described in the LFA by 
\begin{eqnarray}
 F^+(t; \psi, \theta, \phi) & = & \frac{1}{2} \left[ \cos(2\psi) D^+(t; \psi, \theta, \phi, \lambda) \right. \nonumber \\
                              &  - & \left.  \sin(2\psi) D^\times(t; \psi, \theta, \phi, \lambda) \right], \\
 F^\times(t; \psi, \theta, \phi) & = & \frac{1}{2} \left[ \sin(2\psi) D^+(t; \psi, \theta, \phi, \lambda) \right. \nonumber\\
                              & +  & \left.  \cos(2\psi) D^\times(t; \psi, \theta, \phi, \lambda) \right], \\
 \nonumber
 \end{eqnarray}
where $\psi$ is the polarization angle of the wave and the full expressions for the coefficients $D^{+,\times}(t)$ are given in \cite{cornish_rubbo2003}. In general, for a circular monorchromatic binary, we characterize the GW response by the set of seven parameters $\lambda^{\mu}$:  
\beq
\lm =\{ \ln A, \cos i,  \varphi_{0}, \psi, \ln f_{0}, \cos \theta, \phi \}.
\eeq

\subsection{Fourier domain response for galactic binaries}

It is also possible to accelerate the waveform generation by working directly in the Fourier domain. As well as having an acceleration in the waveform generation from not needing to generate long arrays for the time domain
waveforms, nor having to evaluate numerical Fast Fourier Transforms, a further acceleration is gained due to the fact that, for galactic compact binaries the waveform power is packed inside a small number of frequency bins. This procedure has already been derived and tested in other works in the framework of compact galactic binaries \cite{cornish_lisa_2003,timpano_2006,cornish_tests_2007}, so here,  we present only the main concepts behind the formulation.

We can rewrite the expression for the time domain response as
\begin{equation}
s(t) = A_{+}F^{+} \cos( \Phi(t)) + A_{\times}F^{\times} \sin( \Phi(t)),
\label{eq:signalFourier}
\end{equation}
where $A_{+} = A (1 + \cos^{2} (\iota) )$ and $A_{\times} = -2 A \cos (\iota)$. The phase $\Phi$ can be further be written as the sum of three contributions
\begin{equation}
\Phi(t) = 2 \pi f_{0} t + \Phi_{D}(t) + \varphi_{0},
\end{equation}
where we have a binary frequency term, a Doppler modulation term, and the initial phase. To find the analytic expressions for the Fourier coefficients, it is easier to express the 
response in the form 
\begin{eqnarray}
s(t) = \operatorname{Re}\left[ A_{+}F^{+} e^{2 \pi i f_{0} t} e^{i \Phi_{D}(t)} e^{i \varphi_{0}} \right] + \nonumber \\
\operatorname{Im}\left[ A_{\times}F^{\times} e^{2 \pi i f_{0} t} e^{i \Phi_{D}(t)} e^{i \varphi_{0}} \right],
\label{eq:goodform_signal_four}
\end{eqnarray}
where each term inside the brackets needs to be represented in the Fourier domain for an arbitrary observation time $T_{obs}$. 

Dealing with each term in sequence, first of all, given the LFA, the detector functions can be easily expressed as a finite Fourier series of harmonic functions of the modulation frequency $f_m$ 
\begin{equation}
F^{+,\times}(t) = \sum_{k=-4}^{4} p_{k}^{+,\times} e^{2 \pi i f_{m} k t},
\label{eq:detector_harmo_dec}
\end{equation}
where the harmonic coefficients $p_{k}^{+,\times}$ can be deduced from the expressions for $D^{+,\times}$ \cite{cornish_rubbo2003}. Now since the observation time $T_{obs}$ is arbitrary, the associated Fourier coefficients of the detector functions $\tilde{p}_{n}^{+,\times}$ are obtained via the convolution of the above expression with the Fourier transform of a step function, which takes values between $0$ and $T_{obs}$; i.e.,
\begin{equation}
\tilde{p}_{n}^{+,\times}  =  \sum_{k = -4}^{4} p_{k}^{+,\times} \sinc (\pi (k f_{m} T_{obs} - n)) e^{ i\pi  (k f_{m} T_{obs} - n)},
\label{eq:pnFour}
\end{equation}
where we define the cardinal sine function $\sinc(x)=\sin(x)/x$.  In the same way, one can find the Fourier coefficients $\tilde{a}_{n}$ associated with the monochromatic binary frequency term of the phase by a convolution as

\begin{equation}
\tilde{a}_{n} = \sinc (\pi (f_{0} T_{obs} - n)) e^{ i\pi  (f_{0} T_{obs} - n)},
\label{eq:anFour} 
\end{equation}
where the power contribution is peaked around the carrier frequency $f_0$. 

Given the complexity of the Doppler term (due to the cosine in the argument of the exponential), one can express it in a more suitable expression using the Jacobi-Anger expansion
\begin{equation}
e^{i \Phi_{D}(t)} = \sum_{k= - \infty}^{ + \infty}  b_{k}  e^{2\pi i k  f_{m}t},
\end{equation}
where
\begin{equation}
b_{k} =  J_{k}(\beta) e^{i k(\frac{\pi}{2} - \phi)},
\label{eq:bnFour}
\end{equation}
where $\beta=2 \pi f_{0} R_{\oplus} \sin(\theta)$and $J_k(\beta)$ is a Bessel function of the first kind.  This expansion states that the Doppler modulation can be seen as a harmonic function of the modulation frequency. In practice, this sum can be reduced to a finite range by only selecting the significant frequency bins contained within the bandwidth of the signal $B=(1+\beta)$. 

%\begin{equation}
%B = 2(1 + \beta) f_{m} 
%\end{equation}

Once again, a convolution is needed in order to compute the Fourier coefficients associated with $\tilde{b}_{n}$
\begin{equation}
\tilde{b}_{n}  =  \sum_{k = -\infty}^{\infty} b_{k} \sin_{c} (\pi (k f_{m} T_{obs} - n)) e^{ i\pi  (k f_{m} T_{obs} - n)}.
\end{equation}
Putting everything together, we can now write an analytic expression for the Fourier domain response as
\begin{equation}
\tilde{s}_{n}(f) = \dfrac{1}{2} e^{i \varphi_0} \sum_{k} \tilde{a}_{k} \sum_{l} \tilde{b}_{l} \sum_{m} \left( A_{+} \tilde{p}_{m}^{+} + e^{i 3\pi /2} A_{\times} \tilde{p}_{m}^{\times} \right),
\end{equation} 
where $n = k + l + m$. Regarding the ranges of the sum coming from Eqs.~\eqref{eq:pnFour}, \eqref{eq:anFour} and \eqref{eq:bnFour}, only a handful of terms have a significant contribution to the total power of the signal. In fact, by selecting only the terms such that the argument of the cardinal sine is comprised between $\pm18 \pi$, we obtain $99\%$ of the power of the original time domain response \cite{cornish_lisa_2003}.   This puts restrictions on the sums above of $-18\leq k-int(f_0 T_{obs})\leq 18$,
$-(1+\beta)f_m T_{obs}-18 \leq l \leq (1+\beta)f_m T_{obs} + 18$ and $-4f_m T_{obs} - 18\leq m \leq 4f_m T_{obs} + 18$.

A comparison of this method with the computation in the time domain described in Sec.~\ref{Galactic_Waveform}, revealed that the waveform generation computation time can be decreased by a factor of as high as 50. Moreover, the match between the time and Fourier domain waveforms are always higher than 99\%. 

\section{Evolutionary Algorithms}
\label{section_two}
Evolutionary algorithms (EAs) refer to a group of stochastic, population-based algorithms that mimic biological evolution and social behavior
to solve global optimization problems.  An advantage of these algorithms is that they function with no \emph{a priori} knowledge of the problem 
at hand.  This allows, in most cases, a very easy implementation of the algorithm.  Generally, EAs display good
convergence properties and have a small number of control parameters.

Previous studies in GW astronomy have used EAs, but have focused on the implementation of a single algorithm per 
source type~\cite{crowder_lisa_2006,Petiteau:2010zu}.  Our experience has shown that a possible better strategy is to use a combination of algorithms in the development of efficient, 
accurate search and resolution pipelines.    This idea of creating hybrid EAs is not new, and has already been applied in
a number of fields, including GW astronomy~\cite{Omran_2008,gair_cosmic_2009}.

In this work, our goal is to construct a hybrid search algorithm, which predominantly uses a combination of PSO~\cite{kennedy_1995,shi_1998} and DE~\cite{storn_1997}
to iteratively search for monochromatic galactic binaries.  These two methods will be supported by a number of other techniques that will accelerate the convergence
of the algorithm.  In order to both extract the best-fit parameters at the end of the search phase, and conduct a full statistical analysis, we will use a combined Metropolis-Hastings - DE
Markov Chain~\cite{terbraak2006,terbraak2008}.

In this section we describe all algorithms in their base form, before moving on, in the next section, to describing the full construction of
our algorithm.

\subsection{PSO}
\label{PSO_section}
PSO is an evolutionary algorithm that uses swarm dynamics in order to solve optimization problems. This algorithm has already been used in other fields of astrophyscs, such as pulsar timing \cite{taylor_2012,wang_2014}, ground-based GW astronomy \cite{wang_2010} and cosmic microwave background studies\cite{prasad_cosmological_2012}. The PSO is particularly adapted to when trying to find the extremum of multimodal and non-trivial likelihood surfaces. As far as we know, this method has never been tested in the search for galactic binaries using a space based observatory.
	
As mentioned before, the PSO mimics swarm dynamics in nature to find the extremum of a surface. This surface is parametrized by the value of a so-called fitness function that depends on a number of model parameters. In our case, we equate the fitness function to the likelihood $\mathcal{L}(\lm)$. The motion of the 
swarm of particles on the parameter surface is parametrized by a fictitious time parameter $t_{j}$, where $j$ is the identity of the current step. Each particle is then 
evolved
via the standard dynamical PSO equations:
\begin{eqnarray}
X^{i}(t_{j+1}) &=& X^{i}(t_{j}) + V^{i}(t_{j}), \label{Pos_equation} \\
V^{i}(t_{j+1}) &=& w  V^{i}(t_{j}) +  c_{1}  \xi_{1} (P^{i}(t_{j})  - X^{i}(t_{j})) \nonumber\\
&+& c_{2} \xi_{2} ( G(t_{j}) - X^{i}(t_{j})),
\label{eq:Vel_equation}
\end{eqnarray}
where the dynamical variables for particle $i$ are the instantaneous coordinate position $X^{i}(t_{j}) = \{ \lambda^{\mu}_{i} \}$ and velocity $V^{i}(t_{j})$. At each position of the particle on the parameter surface, we associate the corresponding value of the likelihood $\mathcal{L} \left(X^{i}(t_{j})\right)$. 

The velocity of the particle involves a number of quantities specific to PSO.  A further specificity of the PSO algorithm is that each particle retains a partial memory of aspects
of their personal history, while the swarm as a whole retains a sense of global history, both of which effect the velocity evolution of the particle.  With this in mind,
the first important quantities to define are the notions of the personal best, $P_{best}^i$, and group best, $G_{best}$, positions.

We define $P_{best}^{i}(t_{j})$ to be the maximum value of the fitness function for particle $i$ during the duration of its history; i.e., 
\begin{equation}
P_{best}^{i}(t_{j})  = \text{max} \left[ \mathcal{L} \left(X^{i}(t_{k})\right)\text{    , for  } k = 0 \text{ } .. \text{ } j \right].
\end{equation}
The personal best position $P^{i}$ of a particle $i$ at time $t_{j}$ is then defined to be the position where the particle likelihood was equal to $P_{best}^{i}$
\begin{equation}
P^{i}(t_{j}) = X^{i}(t) \text{ if } \mathcal{L} \left(X^{i}(t)\right) = P_{best}^{i}(t_{j}).
\end{equation}
Equivalently, we define $G_{best}(t_{j})$ to be the maximum value of the fitness function of the swarm over its history, or, in other words, the maximum among all $P_{best}^{i}$; i.e.,
\begin{equation}
G_{best}(t_{j})  = \text{max} \left[ P_{best}^{i}(t_{j}) \text{    , for  } i = 0 \text{ } .. \text{ } N_{p} \right],
\end{equation}
where $N_{p}$ is the total number of particles in the swarm. The group best position is then obtained via 
\begin{equation}
G(t_{j}) = P^{i}(t_{j}) \text{ if } P_{best}^{i}(t_{j}) = G_{best}(t_{j}).
\end{equation}
Thus, at any time, each particle of the swarm has a personal memory through $P^{i}_{best}$ and a group memory through $G_{best}$. 

The remaining terms in Eq.~(\ref{eq:Vel_equation}) are an inertia $w$, two acceleration constants $\left(c_{1}, c_{2}\right)$, and two scaling factors $\left(\xi_{1},\xi_{2}\right)$.  For these parameters, the standard values used in the literature are $w = 0.78$, $c_{1} = c_{2} = 1.192$ and $\left(\xi_{1},\xi_{2}\right)\in U[0,1]$~\cite{standard_PSO}. This 
choice of parameters is believed to provide a good compromise between exploration and convergence for the swarm in a reasonable number of steps. However, as 
we will see later, these values can be made to take different values during the evolution of the algorithm.

On further investigation of Eq. (\ref{eq:Vel_equation}), we see that there are three distinct contributions to the evolution of the velocity at each step : \begin{itemize}
\item an inertial term that scales the current velocity with a factor $w$. If $w>1$, the velocity will increase at each step and the particles are more likely to explore the parameter space, while for $w<1$, the particles tend to converge to a single point in parameter space.
\item an acceleration term toward the best position of the particle so far $P^{i}$. This term tends to make the particle explore the neighbourhood of a promising location in parameter space.
\item an acceleration term toward the best position of the swarm so far $G$. This term is shared between all of the velocity equations of the swarm. This is the key term that creates the swarm dynamics: particles will tend to help each other to get to the best position of the group.
\end{itemize}

Since the motion of the particles is governed only by the dynamical equations given above, it is essential to set boundary conditions so that the position of the particles stays within the physical parameter range of interest. One of the common boundary conditions used for PSO is the reflective boundary conditions, where
if the particle crosses the physical boundary,  the 
position of the particle is set at the boundary while the sign of its velocity is reversed; i.e.,
\begin{eqnarray}
X^{i}(t_{j+1}) &=& X_{min} \text{  or  } X_{max}, \nonumber \\
V^{i}(t_{j+1})  &=& - V^{i}(t_{j}) .
\label{Reflective_boundary} 
\end{eqnarray}
Other boundary conditions can be implemented depending on the nature of the model parameter. We give more details on the boundary conditions we have applied in the next section.

Finally, we also set limits on the values that the velocity can take in order to prevent the particle from moving
outside the physical boundary of the problem.  If the velocity computed at time $t_{j+1}$ is superior to a set maximum velocity $V_{max}$, the velocity is then set at this limit; i.e.,
\begin{equation}
V^{i}(t_{j+1}) = 
\left\lbrace 
\begin{array}{ccl}
V_{max}  &  \text{ if} & V^{i}(t_{j+1}) \geq V_{max} \\
- V_{max} & \text{ if} & V^{i}(t_{j+1}) \leq - V_{max} 
\end{array}\right. 
\end{equation}
where $V_{max} = \dfrac{1}{4} (X_{max} - X_{min})$,  $X_{min}$ and $X_{max}$ being the superior and inferior limit, respectively, for the given parameter.

As PSO was originally formulated to mimic flocks of birds, there is no real evolution as the population moves as a whole.  At each point in time, the position
of a particle is influenced by all other members of the swarm.  This is contrast to most EAs.  As we have seen above, the PSO is quite a simple algorithm as it is dependent on only three control parameters $(w, c_1, c_2)$.  The final dependency is the number of particles in the swarm, $N_p$.  However, a downside of PSO is that there is no guide to choosing the optimum value of $N_p$.  As
we will see later, one needs to be careful when choosing $N_p$ in order to achieve a balance between accuracy and runtime of the algorithm.

\subsection{DE}
In this work, DE will be used in two different and distinct manners.  In the first instance, a pure form of DE is used as part of the search phase
of the pipeline.  In the second, it is used as an MCMC variant during the parameter estimation phase.

DE works by evolving a population of $N_p$ candidate solutions in the parameter space.  Each member of the population
evolves via the creation of new candidate solutions by combining existing solutions into a mutant solution.  This
mutation then survives depending on the evaluation of a fitness function (we again use the likelihood function).  A major advantage of DE is that, once again, it
is very capable of handling multimodal solutions.

In this study, we evolve the population via a simplified DE rule : starting with a population of $N_p$ particles at generation $g$, a mutated solution is
produced at generation $g+1$ according to 
\beq
\bar{X}^{i}(g+1) = X^j(g) + \gamma\left[X^k(g)-X^l(g) \right],
\eeq
where $i\neq j\neq k\neq l \in\left[1,..,N_p\right]$.  This means that a minimum of four particles is needed for DE to work.  The differential weight 
$\gamma\in(0,2]$ is a real, constant factor that controls the amplification of the differential vector $\left[X^k(g)-X^l(g) \right]$.  Using this mutant solution,
a crossover solution is created by combining elements of the target solution $X^i(g)$, with the mutated solution $\bar{X}^{i}(g+1)$, to produce a trial
solution $X^{i}(g+1)$.  In general, this solution is then evaluated via a greedy criterion of if $\mathcal{L}(X^{i}(g+1)) > \mathcal{L}(X^{i}(g))$, then accept the
new solution.

In our implementation of the algorithm, we neglect the crossover step, and for reasons that we justify later, we also replace the greedy criterion for
acceptance with an evaluation of the Metropolis ratio,
\beq
H_M = \frac{\mathcal{L}(X^{i}(g+1))}{\mathcal{L}(X^{i}(g))},
\eeq
where the new solution is accepted with a probability of $\alpha = min(1, H_M)$.  As with PSO, the performance of DE is controlled by the values of the
differential weight $\gamma$, the number of particles $N_p$, and the crossover probability.

\subsection{Metropolis-Hastings Markov Chain}
\label{MCMC_section}
The Metropolis-Hastings variant of the Markov Chain Monte Carlo (MCMC) algorithm~\cite{metropolis_1953,hastings_1970} is quite commonly used in the 
field of Bayesian inference.  This quite simple algorithm works as follows : starting with a signal $s(t)$ and some initial template $h(t;\lm)$, we choose a starting point $x(\lm)$ randomly within a region of the parameter space, bounded
by the prior probabilities $\pi(\lm)$, which we assume to contain the true solution.  We then draw from a proposal distribution and propose a jump to another point in the space $x'(\lnu)$.  To compare the performance of both points, we evaluate the Metropolis-Hastings ratio
\begin{equation}
H = \frac{\pi(x')p(s|x')q(x|x')}{\pi(x)p(s|x)q(x'|x)}.
\end{equation}
Here $q(x|x')$ is a transition kernel that defines the jump proposal distribution, and all other quantities are previously defined.  This jump is then accepted with probability $\alpha = \text{min}(1,H)$; otherwise the chain stays at $x(\lm)$.   

In order to improve the overall acceptance rate, the most efficient proposal distribution to use for jumps in the parameter space is, in general, a multivariate Gaussian distribution.   While we have stated previously that the Fisher information matrix (FIM) is not a good choice tool for parameter estimation studies, it is an
acceptable first-order approximation to the true distribution (i.e., the general directionality of the eigenvalues and the scale of the eigenvectors coming from the
FIM produce acceptance rates that allow the algorithm to converge within a reasonable timescale).  We define the FIM as
\beq
\Gamma_{\mu\nu} = \left<\frac{\partial h}{\partial \lm}\left|\frac{\partial h}{\partial \lnu}\right. \right> = -E\left[ \frac{\partial^2 \ln {\mathcal L}}{\partial \lm \partial \lnu}\right],
\label{eqn:FIM}
\eeq
where the second term relates the FIM to the negative expectation value of the Hessian of the log-likelihood, and can be interpreted as a local approximation to the curvature of 
the likelihood surface.  For the multivariate jumps we use a product of normal distributions in each eigendirection of $\Gamma_{\mu\nu}$.  The standard deviation in each eigendirection is given by $\sigma_{\m} = 1/\sqrt{DE_{\m}}$, where $D$ is the dimensionality of the search space (in this case $D=7$), $E_{\m}$ is the 
corresponding eigenvalue of $\Gamma_{\m\n}$ and the factor of $1/\sqrt{D}$ ensures an average jump of $\sim 1 \sigma$.  In this case, the jump in each parameter is given by $\delta \lambda^{\m} = {\mathcal N}(0, 1)\sigma_{\m}$. This type of MCMC algorithm is commonly referred to as a Hessian MCMC.

While the pure form of DE is used during the search phase, we can also use a variant of DE in the parameter estimation phase.  By combining DE
with a Metropolis-Hastings MCMC, commonly known as Differential Evolution Markov Chain (DEMC)~\cite{terbraak2006,terbraak2008}, we can accelerate the convergence
of the MCMC.  In practice, $N_c$ simultaneous chains are used to explore the posterior density.  For chain $i$ at iteration $l$, the next iteration is proposed via
\begin{equation}
X^{i}_{l+1} = X^{i}_{l} + \gamma ( X^{j}_{l} - X^{k}_{l} ),
\label{eq:DE_equation}
\end{equation}
where $i\neq j\neq k\in \left[1,..,N_c\right]$ and the differential weight has the optimized value of $\gamma = 2.38 / \sqrt{2 D}$~\cite{terbraak2006,terbraak2008}, with $D$ being the dimension of the search space parameter.  We note here that we also use this value in the standard implementation of the DE during the search phase.

Again, the number of chains $N_c$ is problem dependent.  While a larger number of chains creates a larger well from which to draw the next proposal, it also
creates a computational bottleneck.  As with previous works~\cite{porter_fisher_2015}, we use a trimmed single chain history to circumvent this problem.  This now means that 
$i\neq j\neq k\in \left[1,..,N_{TH}\right]$, where $N_{TH}$ are the number of trimmed history points.  We have found that $N_{TH} = N_{MCMC}/10$, where
$N_{MCMC}$ is the total number of iterations in the DEMC, provides a good enough history from which to draw points.

\section{Building a hybrid swarm algorithm}
\label{section_three}
In this section, we now present how the previous algorithms can be combined in the framework of the detection and resolution of galactic binaries.

\subsection{Defining the GB parameter space}
In Sec.~\ref{Galactic_Waveform}, we have seen that the gravitational waveform of a circular monochromatic binary is parametrized by a set of seven parameters. This sets the dimensionality of the monochromatic GB problem at $7 \times N$, where $N$ is the total number of ultra compact binary sources in the Galaxy.

In order to reduce the dimensionality of the search space, one can apply an analytical method called the F-statistic \cite{krolac_2004}. The idea is to split the seven 
parameters in two distinct sets labeled extrinsic and intrinsic. The extrinsic parameters, $(i, \psi, A, \phi_{0})$, are related to the detector position and the orientation of the source. The intrinsic parameters, $(f_{0}, \theta, \phi)$, are dependent on the dynamics of the astrophysical source (note that $\theta$ and $\phi$ are intrinsic because of 
the movement of the 
detector with respect to the source). One can then show that for any set of intrinsic parameters, the likelihood can be equated to the F-Statistic in a form that
 automatically maximizes over the extrinsic parameters. Using the F-statistic reduces the dimensionality of the galactic binary search space from $7 \times N$ to $3 \times N$. Full details regarding the F-statistic can be found in Appendix~\ref{F_stat_appendix}.

While the F-Statistic is very useful in finding the maximum of the likelihood surface, it should not be used when mapping the posterior density. In fact, the maximization 
of the extrinsic parameters prevents proper exploration of the posterior density and, as a consequence, parameter estimation can be incorrect \cite{cornish_lisa_2005}. However, unless otherwise stated, we will use the F-Statistic with all algorithms in the search phase of the pipleline. 

\subsection{Setting a detection threshold}
To set a SNR detection threshold for eLISA, we ran a series of null tests.  In this case, we assume that the output of the detector is composed of 
instrumental noise only, i.e. $s(t)=n(t)$.  It has been shown that previous null tests for inspiralling supermassive black hole binaries in eLISA provide
a detection threshold of $\rho=10$~\cite{Huwyler:2014vva}.

In this case we search the detector output using monochromatic waveforms.  To conduct the null test, we used a modified DEMC algorithm.  To encourage the
movement of the chain we use a combination of thermostated and simulated annealing~\cite{cornish_porter_2007_1,cornish_porter_2007_2,cornish_porter_2007_3}.  This replaces the factor of $1/2$ in the likelihood
with an inverse temperature $\gamma=1/(2T)$.  The thermostated annealing, defined by
\begin{equation}
\gamma = \left\{ \begin{array}{ll} \frac{1}{2} & 0\leq \rho\leq \rho_0 \\ \\ \frac{1}{2}\left(\frac{\rho}{\rho_0}\right)^{-2} & \rho > \rho_0  \end{array}\right. ,
\label{eqn:thermann}
\end{equation}
allows the algorithm control over how much heat needs to be injected into the likelihood surface.  In this case, we took $\rho_0=1$ as we wanted the chain to 
begin exploring as quickly as possible.  To extract the threshold SNR, we then need to cool the surface down slowly.  The simulated annealing phase, defined by
\begin{equation}
\gamma = \left\{ \begin{array}{ll} \frac{1}{2}10^{-\xi\left(1-\frac{i}{t_{cool}}\right)} & 0\leq i\leq t_{cool} \\ \\ \frac{1}{2} & i > t_{cool}  \end{array}\right.,
\end{equation}
is then used to cool the surface.  In the above expression, $\xi = log_{10}(T_{th})$, where $T_{th}$ is the temperature at the end of the 
thermostated annealing phase, $i$ is the iteration number, and $t_{cool}$ is the cooling schedule for the simulated annealing.  For the null tests, we used
$5\times10^4$ iterations each of thermostated annealing, simulated annealing, and standard DEMC.

We ran the above algorithm 50 times, with different starting configurations.  In each case, the algorithms returned ``detections" with SNRs of $5.9\leq\rho\leq8.7$.
To account for the possibility of higher values from different noise realizations, we thus decided to set the detection threshold at $\rho=9$.

\subsection{Single-source detection}
Our initial goal was to investigate and define regions of functionality for each component of the search pipeline.  This top-down approach would allow us to 
confidently detect a source within a small region of parameter space surrounding the binary, and then expand the size of the search space, solving problems
as we go.  With this in mind, we defined a search region around the binary in each case by $\lambda^i\pm n\Delta\lambda^i$, where $n>1$.  In our initial investigations, we
defined $\Delta\lambda^i=\sigma_i$, where $\sigma_i$ is the standard deviation calculated using the FIM (we point out again that we are not using the FIM
for parameter estimation purposes, but only to define a region of search space).  In later investigations where we are dealing with a wider search space, for frequency, we take $\Delta f_0=f_m = 1/yr$.

We used two criteria to assess whether the algorithm was working for a given width $n$: does it recover the SNR of the source, and are the recovered parameters less than $3 \sigma$ away from the true solution? If those two conditions were met for repeated simulations, we increased the width of the search space. Thus, we were able 
to control both how our algorithm works and observe exactly when it breaks down. 

In order to test the performance of the algorithms, we started with a simple search where the data contained a single binary and an observation time of one year.  Two different
sources were selected for this first test phase. The first one was a fiducial low-frequency source, which allowed us to reduce runtime of the algorithm during development and solve
problems in our initial investigations quickly.   The parameters for this binary were $\iota=1.74, \phi_0=1.664, A=1.2276\times10^{-20}, \psi=1.884, f_0=5.2234\times10^{-4}\,\text{Hz}, \theta=1.273$, and $\phi=1.8845$, where the angular values are in radians.  The signal had an SNR of $\rho\approx54$.  The second source was the WD-WD GB, RXJ0806.3+1527. This 
source is one of the main verification binary candidates for eLISA and should be the one recovered with the highest SNR \cite{kilic_ultra-compact_2013}. The parameters for this 
binary are $\iota=0.663, \phi_0=5.857, A=6.378\times10^{-23}, \psi=1.741, f_0=6.2203\times10^{-3}\,\text{Hz}, \theta=1.162$, and $\phi=3.612$~\cite{roelofs_2010}.  For 1 year of data, the SNR for
RXJ is $\rho\approx25$

\subsubsection{Initial search with PSO}
\label{section2_3}
The initial step in the construction of our algorithm was to implement the PSO to see how well it performs, on its own,  in the detection of galactic binaries. There were two motivations to use PSO as a baseline algorithm :
\begin{itemize}
\item In comparison with MCMC based algorithms, the performances of PSO seemed to be more efficient in finding the maximum of complicated posteriors \cite{prasad_cosmological_2012}.
\item The swarm-like feature of the algorithm, such as the personal and group best position, seemed to be powerful and could easily be adapted, or
complemented,  in other situations.  
\end{itemize}

Since the PSO algorithm described in Sec.~\ref{PSO_section} is very general, we need to explain some specificities for the GB search. The set of intrinsic parameters used for a single source PSO search is $\lambda^{\mu} = \{ \ln f_{0}, \theta, \phi \}$. We do not use the cosine of the colatitude in this case, as we want the particles to evolve freely on the 2-sphere of the sky. In terms of boundary conditions for the angles, we let the particles evolve freely in absolute values of the sky angles.  If the value
of the sky position goes outside the physical boundary, we coherently remap the source into the natural boundaries $\theta\in[0,\pi]$ and $\phi\in[0,2\pi]$. For the frequency, we used the reflective boundary condition given in Eq.\eqref{Reflective_boundary}.

For the first application of PSO, we focused solely on the low frequency source.  We initially used 10 particles with standard control parameter values of $w = 0.72$, 
$c_{1} = c_{2} = 1.192$ for a total number of steps of $500$. Using $\Delta\lambda^i=\sigma_i$, for every 
chosen value of width $n$, we launched ten simulations with different initial conditions. In this configuration, the PSO works well up to $n=10^2$, where one of the 
simulations did not converge to the true solution. This value of $n$ corresponds to a $\sim3 f_{m}$ frequency bandwidth, where some of the secondary peaks in the likelihood, caused by the Doppler modulation, are now accessible to the particles. In the failed simulation, the swarm was not able to escape a secondary mode in the likelihood surface. 

One possible solution for this problem is to increase the number of particles, $N_{p}$, thus giving the swarm more opportunities to find the main mode. This solution is not costless since it also increases the total computation time. This is why we decided to keep this as a last-resort solution, and instead introduce a new control on the swarm. In order to fine-tune the 
behavior of the swarm we now allow the constant inertia $w$ to vary over time, i.e. $w=w(t_j)$.  In the dynamical equations, a value of inertia superior to 1 increases the exploration of the swarm, while convergence of the swarm is improved for inertias inferior to 1. A good compromise between exploration and convergence can be found by introducing a type of ``inertia annealing". This method is not new and has already applied in other works, albeit in a different form~\cite{wang_2014}.   Starting with an initial inertia $w_i$, we
cool the inertia according to 
\begin{equation}
w(i) = 
\left\lbrace 
\begin{array}{lcl}
w_{f}  10^{ \log_{10}(\frac{w_{i}}{w_{f}}) (1 - \frac{i}{T_{w}} )}  &  \text{ if} & 0 \leq i \leq T_{w} \\
 & & \\
w_{f} & \text{ if} & i > T_{w} ,
\end{array}\right. 
\end{equation}
where $w_i=1.2$, the final inertia is $w_f=0.78$ and $T_w=500$ is the cooling time (which we take to be equal to the total number of steps in the algorithm).  With this 
new annealed version of the algorithm,  the swarms now successfully found the source up to a value of $n=10^4$. For this value of
 $n$, the search space now covers the full sky and is $\sim300 f_{m}$ wide in frequency.

 Beyond this value of $n$, we seemed to come to a natural limit in
 the PSO algorithm.  The efficiency of the PSO is based on the ability of members of the swarm to find good positions in parameter space,  and drag the whole set of particles toward them. One of the fundamental requirements is then to have a good exploration of the parameter space so that the swarm is not confined in some small part of the 
 likelihood surface. For medium size widths such as the one with $n=10^4$, we noticed that for some initial conditions, the swarm was not able to explore a large 
 enough space to find the area of interests. It is known that PSO can very quickly converge to secondary maxima, especially on a complicated surface.  Once there, it becomes very difficult to move the algorithm on to a better solution.  This is why we decided to introduce a DE step that would provide the 
 swarm a greater ability to explore the parameter space.

\subsubsection{Combining PSO with DE}
In the DE part of the combined PSO-DE algorithm, all of the particles are evolved sequentially using Eq.~(\ref{eq:DE_equation}). This is considered to be one step in the pipeline. At
 the end of the step, we update the values of the personal and group best positions of the swarm.   Because of its different nature, DE is more suited for global exploration and
  large moves in the search space. Therefore, the main idea was to alternate between standard PSO blocks as described before and DE blocks that would help find better $P^{i}$ and $G$ for the next stage of PSO.   From the beginning, we also decided to use the thermostated and simulated  annealing schemes for the DE. 
  
We implemented our algorithm using two series of PSO and DE blocks of 125 steps each for a total number of 500 steps.  As we have already described, both the
PSO and the DE have their own specific annealing schemes.  As we alternate between the PSO and the DE, we also need to alternate between the annealing
programs.  To ensure that the use of these schemes was optimal, they were implemented in the following manner :
both the inertia annealing (for the PSO) and the thermostated/simulated annealing (for the DE) were run simultaneously.  This ensured that each annealing scheme
cooled over a long period.  It also meant that when one scheme was in use, the other was running in the background, ensuring that when it came back into use, it 
would be at the same level as if it was being applied to a single algorithm pipeline only.  So, for the PSO, we set $T_{w} = 500$. This means that at the end of the algorithm, the inertia is equal to the final value of 
$w_{f} = 0.78$, even though the actual total number PSO steps is 250. During the DE phase, we used a thermostated annealing scheme during the first 125 steps, with a threshold of $\rho_0 = 5$ to encourage movement in the parameter space.   We then set $T_c=375$ for the simulated annealing phase, meaning that at the end
of the pipeline, the heat is unity.  In this new combined configuration, the algorithm managed to always find the source for $n=10^5$,  which corresponds to a frequency bandwidth of $\sim3000 f_{m} \approx 10^{-4} \text{Hz} $ 

At this point, we chose to test the algorithm on a full $1 \text{mHz}$ frequency band to observe its performance. The interest here is to see whether, given the width of the signal
in comparison to the width of the search space, the algorithm would find the source, or repeatedly get stuck on strong peaks in the noise, i.e., a kind of ``needle in an empty 
field" problem.  Since the frequency band is now much wider, we 
decided to keep the structure of the previous algorithm, but it quickly became clear that we finally needed to increase the number of particles in the swarm.  

We thus 
changed the number of particles from $15$ to $40$ to facilitate a wider exploration (we comment further on the number particles required below).   In this 
configuration, the algorithm was doing quite well, but was not able to satisfy the two convergence criteria for all the simulations. 

In fact, we observed two 
main weaknesses in the movement of the swarm: in the case where the source has not yet been detected, the group best position $G$ may be lying on a distant noise
peak.  A particular particle may land in the vicinity of the source, and in so doing, improve its personal best position $P^i$.  However, as $ {\mathcal L}(P^i) < {\mathcal L}(G)$, the group best position will never move into the area of interest.  Furthermore, while $P^i$ stays close to the source, the particle itself wanders off to explore the rest
of the parameter space.  If a new personal best position is found, it can actually pull $P^i$ away from the source, delaying detection.  In the second case, a source can be
detected (i.e., its SNR beats the threshold), but the group best position ends up on a secondary solution.  In order to improve on this current solution, we have to wait until
one of the particles lands at random on a better solution and thus improving $ {\mathcal L}(G)$.

Upon investigation, the culprit in both cases was the width of the search window.  A wide search window allows for a very diffuse swarm.  This means that any chance of a
local exploration is dramatically reduced.  In this case, particles will visit the region of parameter space close to the source once, and then maybe never again after.  One solution would be to overpopulate the search space with particles, increasing the possibility that certain regions would be visited by many particles, many times, during the
runtime of the algorithm.  However, this would lead to a very slow algorithm.  So, in order to solve these two issues, while keeping the number of particles small, we introduced two further features to the algorithm that enforce a more local exploration and reduces the width of the search space.

\subsubsection{Uphill Climber}

The motivation for this type of move is fact that the swarm algorithm's lack of local exploration possibilities. In the previous test cases, the frequency band was small enough that the swarm was able to locally explore good positions on the likelihood surface. For wider frequency bands, the swarm is more diffuse, and the PSO-DE 
blocks do not provide an opportunity for the swarm to locally explore the personal best positions $P^{i}$. 

The idea was then to implement a new block of MCMC-like movements that would be related, not to the individual swarm particles, but directly to the $P^{i}$. Thus, at 
some time $t_j$, we stop the global movement of the swarm and locally explore all $P^i$ positions that have been spotted so far with the PSO and DE blocks.

The uphill climber (UC) scheme was first introduced in \cite{gair_cosmic_2009} as a greedy criterion proposal, i.e., if the new solution has a higher fitness than the starting solution,
move there immediately.   In the current context, this move is implemented as follows : for each $P^i(t_j)$, and as we are using the F-Statistic, calculate the
FIM on the 3D-projected subspace at that point, i.e., iteratively project the $D$-dimensional FIM onto a $D-1$ subspace according to 
\begin{equation}
\Gamma_{\mu \nu}^{(D-1)} = \Gamma_{\mu \nu}^{(D)} - \dfrac{\Gamma_{\mu \kappa}^{(D)} \Gamma_{\nu \kappa}^{(D)}}{\Gamma_{\kappa\kappa}^{(D)}}.
\end{equation}
As with the MCMC algorithm, use the eigenvalues and eigenvectors of the projected matrix to construct jump proposals, and accept or reject according to the 
greedy criterion ${\mathcal L}(\lambda^{new}) > {\mathcal L}(\lambda^{old})$.

In order for the UC to be effective, it needs to be implemented a number of times successively as the acceptance rate is usually quite low ($<10\%$).  Although
low, and probably coming from the fact that we do not update the eigenvectors and eigenvalues after a move has been accepted, if the UC is used $N_{UC}$ times
in sequence (with $N_{UC}\geq 10^2$), it can easily move the position of $P^i$ between $10-30$ $f_m$ in frequency.  This greatly accelerates the convergence
of the algorithm.  To make the UC fit into the overall structure of the pipeline, and especially with the annealing schemes, we count the $N_{UC}$ iterations as one
step in the algorithm.  This means that we take the thermostated/simulated annealing temperature at $t_j$ and keep it constant during the UC moves.

\begin{table*}[t]
\begin{tabular}{|l|c|c|c|c|c|c|c|c|}
 \cline{2-9}
\multicolumn{1}{c|}{} & $\iota$ & $\phi_{0}$ &  $A$ & $\psi$ & $f_{0}/mHz$ & $\theta$ & $\phi$ & SNR  \\
  \hline
Source 1 & 1.74 & 1.664 &  $1.2275\times10^{-20}$ & 1.884 & 0.52234 & 1.273 & 1.8845 & 53.92 \\
 &1.736 & 1.707 & $1.2429\times10^{-20}$ & 0.3076 & 0.52233969 & 1.239 & 1.9097 & 53.95\\
  \hline
RXJ0806.3 &  0.663 & 5.8565 & $6.378\times10^{-23}$  & 1.74 & 6.2202766 & 1.1624 & 3.6116 & 24.93 \\
 & 1.143 & 1.785 & $9.769 \times10^{-23}$ & 0.742 & 6.2202745 & 1.1600 & 3.6236 & 25.53\\
  \hline
\end{tabular}
\caption{The injected and recovered parameter values for our two test sources. The values for RXJ0806.3+1527 are taken from~\cite{roelofs_2010}.  We provide
the recovered results for two simulations that end up on the response invariant symmetric solution $(\phi_0, \psi)\rightarrow(\phi_0\pm\pi, \psi\pm\pi/2)$ .}
\label{tab:Source_Param_Table}
\end{table*}

\subsubsection{Swarm Strengthening and Culling}

The second feature was introduced due to the fact that the likelihood surface has many secondary peaks because of the multi-modal nature of the detector response.  One of the strongest secondary
solutions comes from the Doppler modulation of the phase. For some simulations, the swarm was able to get close to the source but was stuck on one of these Doppler induced peaks. As mentioned above, when the frequency band is large, even with a local exploration of the likelihood surface, only a handful of particles explore the area 
of interest, and those that do so may converge to one of the secondary solutions.   In order to prevent this situation, we decided to add a second phase in the overall algorithm where we strengthen a good solution by moving half of the swarm to an interval of frequency around $G$ given by $[f_{0}(G) \pm 10 f_{m}]$. The size of this band is motivated by the assumption that, if this feature is used late enough in the pipeline, the principal mode should not be further away than a few $f_{m}$ from the secondary solutions. 

Furthermore, while it is clear that an increased number of particles is necessary in the early stages of the pipeline to ensure a wide exploration of the parameter
space, later on, some of these particles can converge to low fitness regions of the likelihood surface.  Seeing as they never evolve to the region of 
interest during the runtime of the pipeline, we cull the 50\% of the swarm that is not drawn to the region around $G$.  This also helps speed up the runtime
of the algorithm in the second phase of the pipeline.

\subsubsection{Single source search over a 1-mHz band}
\label{section_three_finalsearch}
As the primary goal of this study is accuracy, rather than speed, the final version of our algorithm is as follows : the pipeline is now 1250 steps long, beginning
 with 40 particles. In the first phase, the structure is two blocks of 150 steps each of PSO and DE, followed by 200 steps of UC. For the PSO we use an inertia annealing 
 scheme with $T_{w} = 750$, while for DE and UC, we use a thermostated annealing scheme. At the end of this phase, we move the swarm that is now 20 particles in size
 to the vicinity of $G$ and kill the other 20 particles.  In the second phase of the pipeline, we have a structure of 75 steps of PS0, followed by 75 steps of DE and 100 steps of UC. For DE 
 and UC, we use a simulated annealing scheme with an initial temperature equal to the temperature at the end of the first phase and a cooling time of $T_{c} = 250$. 
 We again point out that this structure is not optimized and will be a focus of our research in the future.

The search algorithm was then tested for the detection of the low frequency source, using a 1-mHz search band. We ran 10 simulations with different initial conditions, 
and in each case recovered the source according to both detection criteria.  In Table~\ref{tab:Source_Param_Table}, we present both the injected and recovered parameter values
from the search.   For this source, the total runtime for the search algorithm was ~1 minute on an Intel Xeon 2.6 GHz processor.

\begin{figure*}[t]
\begin{center}
\epsfig{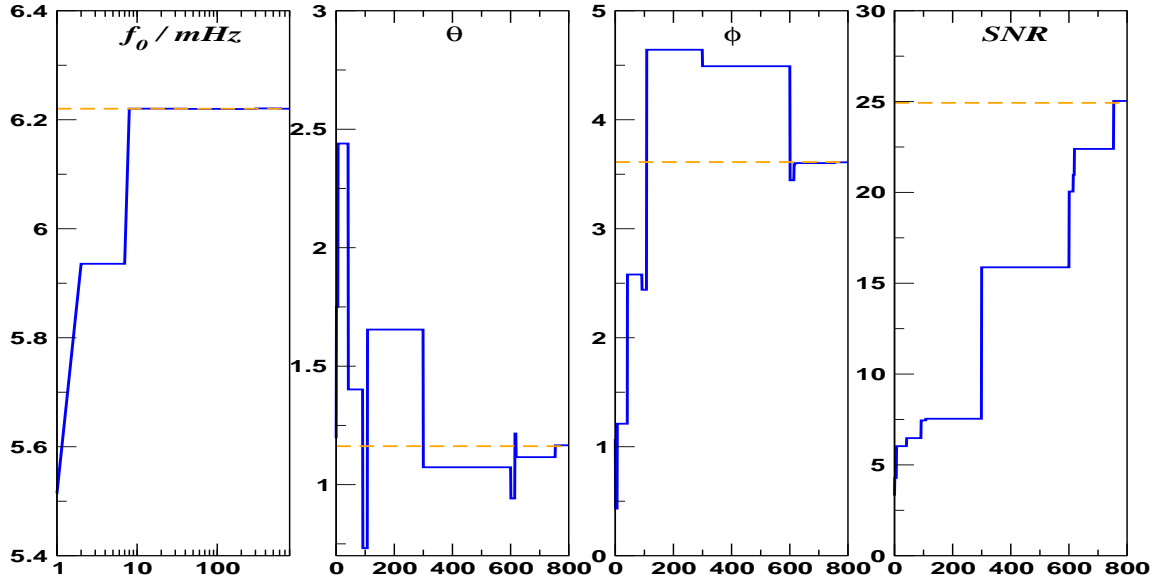}
\end{center}
\caption{A plot of the movement of the group best position $G$ during the search for RXJ0806.3+1527, for the parameters $(f_0, \theta, \phi)$ as a function of the
number of steps in the algorithm. In each cell the 
true values are represented by the dashed (orange) lines.  We can see that $G$ converges to the true frequency in less than ten steps, whereas the sky angles take
longer to find.  In the final cell we plot the evolution of SNR.  Here the large jumps in SNR at 300, 600 and 750 steps are due to the uphill climber move.} 
\label{fig:rxjsearch}
\end{figure*}

We then ran the algorithm on the more realistic source, namely the WD-WD verification binary RXJ0806.3+1527. Again, in all simulations, the algorithm was able to find
 the global solution.  In Fig~\ref{fig:rxjsearch} we plot the evolution of the group best position $G$  for the parameters $(A, \iota, f_0)$, as well as the evolution of SNR, as a 
 function of step number.  We can see that the algorithm performs very well and we converge close to the true value of frequency in less than 10 steps of the algorithm.  The
 sky angles take a lot longer to pin down, but are also eventually found.  In the final cell of the figure we plot the evolution of SNR as a function of step number.  We draw
 attention at this point to the values of the parameters and the SNR at 300, 600 and 750 steps in the plots.  At each of these values we have an uphill climber move.  We can see from looking at $\theta$ and $\phi$
 that the UC moves are responsible for locking $G$ onto the true values of the parameters.  In the SNR plot, we can see that the UC produces large jumps in the
 recovered SNR.  This again justifies the inclusion of this type of move.  For this source, and due to the much higher sampling rates required, the total time for the search algorithm  was $12.5$ minutes.  Again, we point out that there is a large potential for future optimization of the algorithm.  We should also highlight here that for long periods
 of time in the plot, it looks like the algorithm has failed to move the solution and that it is static.  We remind the reader again that the quantity being plotted is the group best value $G$.  An investigation
 of the particle positions displays a constant evolution of the swarm.
 
 In Table~\ref{tab:Source_Param_Table} we again present both the injected and recovered parameter values  from the search.  We should point out that the 
recovered results presented are for two simulations that end up on a $(\phi_0, \psi)\rightarrow(\phi_0\pm\pi, \psi\pm\pi/2)$ symmetric solution that leaves the response $s(t)$ 
invariant.  This seems to be
quite a common feature for a single channel observatory, and is a perfectly valid solution due to the multimodal nature of the response.  In keeping with recent articles on supermassive black hole binaries~\cite{porter_fisher_2015, porter_alone_2015}, we also
carried out a full statistical analysis on the Markov chains for each source.  For the low frequency source, all distributions had low values of skewness and kurtosis.  However,
for RXJ0806.3+1527, the posterior distributions were highly skewed and/or had a high kurtosis for a number of parameters.  This demonstrates that the FIM should also 
be avoided as a parameter estimation tool for compact galactic binaries.  

As a consequence, in our full analysis, we present $99\%$ credible intervals, ${\mathcal C}$, for the parameter errors such that
\beq
\int_{\mathcal C} p\left(\lm | s\right) d\lm = 1-\alpha,
\eeq
where for a $99\%$ BCI, $\alpha = 0.01$.  This is a degree-of-belief statement that the probability of the true parameter value lying
within the credible interval is $99\%$; i.e.,
\beq
{\mathbb P}\left(\lm_{true} \in {\mathcal C} | s \right) = 0.99.
\eeq
For the positional resolution of the source, we can define an error box in the sky according to~\cite{cutler_1998} 
\begin{equation}
 \Delta\Omega = 2\pi \sqrt{\Sigma^{\theta\theta}\Sigma^{\phi\phi}-\left(\Sigma^{\theta\phi}\right)^{2}},
\end{equation}
where 
\begin{eqnarray}
 \Sigma^{\theta\theta} &=& \left<\Delta\cos\theta\Delta\cos\theta\right>,\\
  \Sigma^{\phi\phi} &=& \left<\Delta\phi\Delta\phi\right>,\\
   \Sigma^{\theta\phi} &=& \left<\Delta\cos\theta\Delta\phi\right>,
\end{eqnarray}
and $\Sigma^{\m\n} = \left<\Delta\lambda^{\m}\Delta\lambda^{\n}\right>$ are elements of the variance-covariance matrix,  calculated directly from the DEMC chains themselves.  As well
as the sky error box, we also calculate the orthodromic distance between the true and recovered sky positions.  This is given by the Vincenty formula
\begin{widetext}
\begin{equation}
\Delta \sigma = \text{arctan} \left( \dfrac{\sqrt{\left(\cos \phi_{R} \sin \Delta \theta_{L}\right)^{2} + \left( \cos \phi_{T} \sin \phi_{R} - \sin \phi_{T}  \cos \phi_{R}  \cos \Delta \theta_{L} \right)^{2}}  }{\sin \phi_{T}  \sin \phi_{R}  + \cos  \phi_{T} \cos \phi_{R}  \cos \Delta \theta_{L}  }    \right),
\end{equation}
\end{widetext}
where we define the true longitude $\phi_{T}$ and latitude $\theta_{T} $ of the source, the recovered longitude $\phi_{R}$ and latitude $\theta_{R}$, and  $\Delta \theta_{L} = \theta^{T}_{L} - \theta^{R}_{L}$, and the value of $\Delta\sigma$ is in radians.

As the RXJ source is one of the main verification binaries of a space-based mission, we should comment here that for the eLISA configuration, with 1 year of data, the 
recovered frequency has an absolute error of 2 nHz or $0.06\,f_m$, corresponding to a percentage error of $\sim10^{-5}\%$. The recovered amplitude has an absolute error of $\sim3\times10^{-23}$, while the absolute error in the measurement of inclination is approximately 29 degrees. And even though the sky position is known from EM observations, the sky error box based on 1 year of observation is $\Delta\Omega \approx0.6\,\deg^2$.

We should take some time here to talk about both the number of particles in the swarm, $N_p$, and the stopping criterion used for the pipeline.  For the algorithm described above, we initially used 40 particles before culling the 
swarm to 20.  In order to test for accelerated convergence, we ran the algorithm with different numbers of initial particles, up to $N_p=10^2$.  In no circumstance did
we observe an acceleration of convergence that would convince us start with more than 40 particles.  As expected, however, we did see an increase in the 
runtime of the pipeline that came without a corresponding increase in convergence or accuracy.  As with most search algorithms, there is no predefined stopping
criterion.  This case is identical.  An investigation of quantities such as $P^i$ or $G$ did not suggest any obvious way of using these parameters as
stopping criteria.  In this particular study, as we know the true values \emph{a priori}, the number of steps in the algorithm was chosen to ensure that we always found the true
source.  We intend to investigate a more rigorous stopping criterion in a future work.

\begin{figure*}[t]
\begin{center}
\epsfig{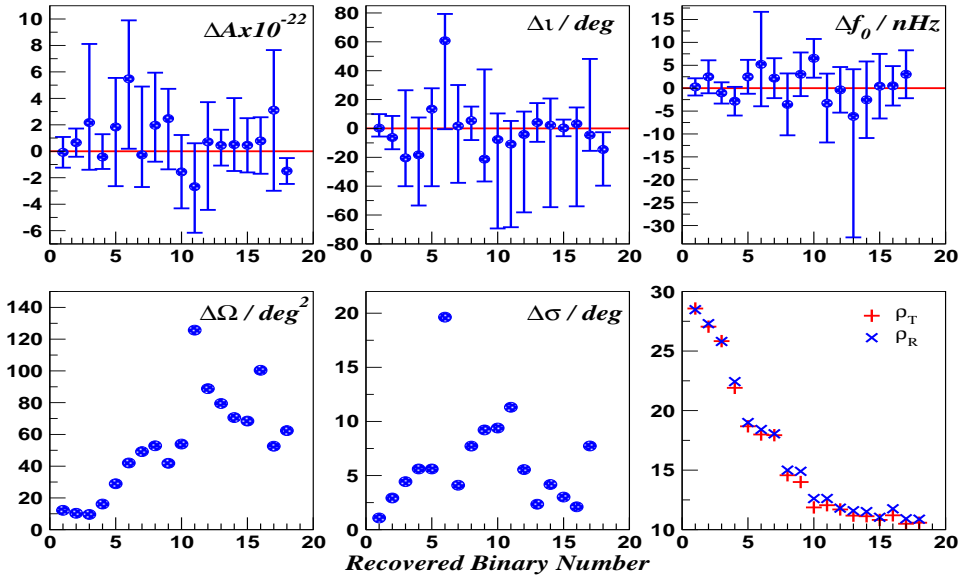}
\caption{Results obtained from the recovered sources in data set 1. In the top row, we present the $99 \%$ credible intervals and (true subtracted) median values for amplitude A (top left), inclination (top middle) and frequency (top right). In the bottom row, we present the values of the sky error boxes, the orthodromic distance, and the values of the true and recovered signal-to-noise ratios.}
\label{fig:Set1_Confidence}
\end{center}
\end{figure*}

\section{Analysis and Results}
\label{section_four}

\subsection{Presentation of the problem}

After the single source search, we decided to test our algorithm in a situation closer to reality where the data set contains several binaries. We have built two data sets with different attributes

\begin{itemize}
\item The first set of data contains 18 sources in a frequency band of 1-mHz width from $0.5$ to $1.5$ mHz. The SNR of the sources have been taken between $10.5$ and $28$. The minimum distance in frequency between two sources is $273 f_{m}$, meaning that there is no confusion between the sources, but there is plenty of opportunity
for the algorithm to get stuck on a strong peak in the noise.
\item The second data set contains 30 sources in a frequency band of $10^{3} f_{m} \approx 30 \mu Hz $, centered at 1 mHz. The SNRs of the sources have been taken between $10.9$ and $35$. The minimum distance in frequency between two sources is $9 f_{m}$, meaning that there is now a mild confusion between the sources.
\end{itemize}

At the end of the pipeline described above, we are faced with two practicalities that require different treatments.  The first is that we need to have a way of quickly and
accurately subtracting the recovered source before moving onto the search for the next source.  The other is a full statistical analysis of the recovered source.

To complete the first task, at the end of the search phase, we run a $4\times10^4$ iteration DEMC with a constant inverse temperature of $\gamma = 1/2$.  This gives the 
chain a chance to become statistically independent.  We then run a further $2\times10^4$ iterations where we superfreeze the chain to extract the Maximum Likelihood
Estimator (MLE)~\cite{cornish_lisa_2005}.  This is done by using a simulated annealing phase, where we use an inverse temperature of $\gamma = 1/(2T)$, starting with $T_i=1$ and ending
with $T_f=10^{-5}$.  At the same time, to conduct the statistical analysis and calculate the credible intervals, we run a $10^6$ iteration DEMC, where we neglect the 
first $2\times10^4$ iterations as ``burn-in".

\subsection{Data Set 1}

For the first data set with no confusion, we detected 17 sources at the main frequency peak and 1 source on a secondary shifted frequency peak.  In 
Fig.~\ref{fig:Set1_Confidence}, we plot the (true subtracted) recovered median values for the parameters $(A,\iota,f_0)$, i.e. $\Delta\lambda=\tilde{\lambda}_R-\lambda_T$, as well as the 99\% credible 
intervals, as a function of the recovered binary.  We also plot the sky error box $\Delta\Omega$, the orthodromic distance between the true and recovered sky positions $\Delta\sigma$, as well as the SNR.

For the amplitude and inclination, all the injected source values are contained in the credible intervals, except for source 6. For this source, the binary has almost no inclination.  This leads to an extremely high anticorrelation between the parameters of -0.949.  An inspection of the correlation matrices for the binaries using the information from the 
chains, we noticed that a high correlation value between amplitude and inclination leads to larger credible intervals. For this data set, we found nine binaries with correlations 
between $A$ and $\iota$ superior to $0.9$ in absolute value. In addition, most of the credible intervals for inclination are not symmetric around the recovered values, due to the 
high skewness and/or kurtosis of the posterior distributions. This confirms that credible intervals are indeed important to use in this situation. In terms of frequency, all the 
frequencies are recovered in the interval except for binary number 10.  A closer inspection reveals that the median frequency from the chain is at $3.95 \sigma$ away from true frequency, where $\sigma$ is the standard frequency deviation from the chain. However, the SNR of the recovered signal is the same as the source and the residual is below the noise. This means that we indeed have a detection. In the majority of cases, the recovered median frequencies were within 7 nHz, or $0.2\, f_m$ of the true values.  We should point out that source 18 is not represented in the frequency cell.  We will tackle this binary separately below.

If we now focus on the bottom row of Fig.~\ref{fig:Set1_Confidence}, we can see that, as expected, the sky error box tends to grow as a function of diminishing SNR.  In almost
all cases, the sky error box is smaller than 100$\deg^2$, with some of the high SNR sources having error boxes of $\sim10\deg^2$.  One binary (source 11) has a large error
box ($\sim150\deg^2$).  On inspection, we observed a large imprecision in the resolution of colatitude for this source, but the recovered value was
within a $3\sigma$ of the true value.  From the orthodromic distance, we can see that the recovered sky solutions are usually within $~\sim10\deg$ of the true value.  In the cell representing the SNRs, it is interesting to see that most of the binaries have been found in order of SNR. This essentially means that our search algorithm is able to locate the brightest signal on a 1-mHz band and finds the maximum of the fitness function in a noisy background.

The final source (\#18), where we ended up on a secondary Doppler peak, was the binary with the lowest injected SNR of the set ($\rho_{T} = 10.58$).  In all of our simulations,  the MLE 
frequency extracted at the end of the search phase corresponded to a frequency that was $\sim 3 f_{m}$ away from the true value. However, the recovered SNR from the MLE ($\rho_{R} = 11.08$) was actually 
higher than the injected value due to noise. It has been pointed out in the literature that this situation can happen, especially for sources with SNR close to the threshold \cite{crowder_lisa_2006}.   As a sanity check, we ran 10 simulations with different noise realizations and tried to find this single source. For each
 simulation we managed to find the source on the true frequency peak, confirming that it was indeed a problem of noise realization, and not a fundamental problem with 
 the algorithm.

To further test how we had done, we also computed the overlap between the true and recovered signals, where the overlap between $h_{T}$ and $h_{R}$ is given by
\begin{equation}
\mathcal{O} = \dfrac{\left<h_{T}\left|h_{R}\right.\right>}{\sqrt{\left<h_{T}\left|h_{T}\right.\right> \left<h_{R}\left|h_{R}\right.\right>}}. 
\end{equation}
The 17 fully recovered sources had overlaps above 0.9, with values as high as $0.999$ for the highest SNR source.  For source 18, the overlap was 0.82.

\begin{figure}
\includegraphics[scale=0.4]{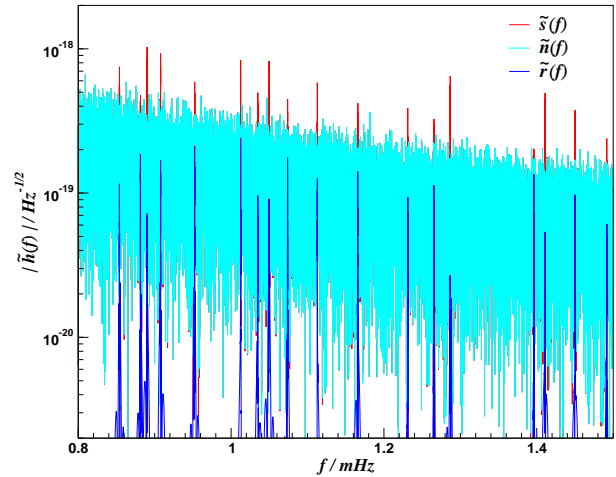}
\caption{A plot of the power spectra for the injected data set, the instrumental noise and the residual for data set 1.}
\label{fig:Set1_Residual}
\end{figure}
We can see what this means graphically by plotting the residual of the subtracted data set against the original signal and the instrumental noise.  The residual power is
an indication of how much we have disturbed the data set through the imperfect subtraction of a source.  We define the residual as
\begin{equation}
\tilde{r}(f) = \tilde{n}(f) - \left( \tilde{s}(f)- \sum_{i=1}^{N_{s}}  \tilde{h}^{MLE}_i(f)\right).
\end{equation}
with $N_{s}$ the number of recovered sources.  In Fig~\ref{fig:Set1_Residual}, we plot the power spectra for the total signal with noise $\tilde{s}(f)$, the instrumental noise only $\tilde{n}(f)$ and the residual $\tilde{r}(f)$.   If the residual is below the level of the noise, then the subtraction process has been sucessful. For the first data set, we can see that the level of the residual is always below the level of noise even for the binary where the recovered frequency was $3 f_{m}$ away. 

We also carried out one final extra check to test the performance of the subtraction process. We ran 40 simulations, using different initial configurations, on the binary
subtracted residual data set.   If we had not properly extracted all binaries, we should in this case ``detect" another source.  However, all simulations ended up with a 
``detection" with an SNR inferior to the SNR threshold. Moreover, as some of the returned solutions were clustered around specific frequencies, we inspected these
value and confirmed that they were pure noise peaks at more than $50 f_{m}$ away from any injected frequency signals.

\subsection{Data set 2}

\begin{figure*}
\begin{center}
\epsfig{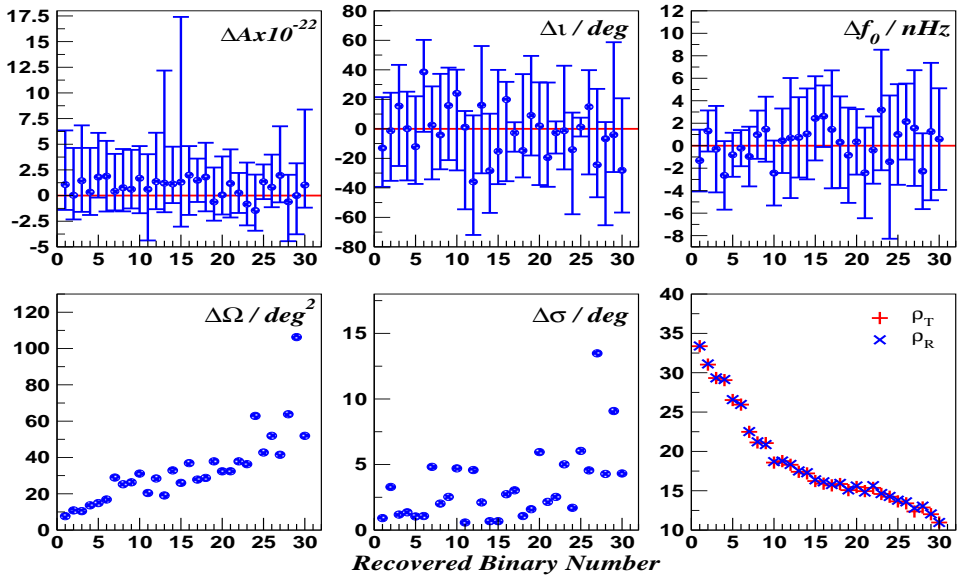}
\caption{Results obtained from the recovered sources in data set 2. In the top row, we present the $99 \%$ credible intervals and (true subtracted) median values for amplitude A (top left), inclination (top middle) and frequency (top right). In the bottom row, we present the values of the sky error boxes, the orthodromic distance and the values of the true and recovered SNRs.}
\label{fig:Set2_Confidence}
\end{center}
\end{figure*}

For the second data set, the algorithm managed to find all sources on the $10^{3} f_{m} \approx 30 \mu \text{Hz}$ band.   
In Fig.~\ref{fig:Set2_Confidence}, we again plot the credible intervals, sky errors, and SNR.  For this data set, all the true values lie in the $99 \%$ credible intervals for amplitude, inclination, and frequency. Once again, we observe that the widths of the credible intervals for $\iota$ and $A$ depend on the values of these parameters and their correlation.  For instance, the largest credible interval for amplitude (binary 15) corresponds also to a high correlation between amplitude and inclination of $0.951$ and a high asymmetry in the posterior density. In terms of frequency, the recovered values are very good and all lie at less than $3$ nHz or $0.1\,f_m$ from true frequency.

For the sky positions, the sky error boxes have expected values.  For this data set, the orthodromic distance between injected and recovered sky positions is lower, with
nearly all sources recovered within $5\deg$ of the true value . Once again, the size of the sky error box increases as the SNR decreases, and the binaries
we again found in order of decreasing SNR.  In this case, and probably due to the smaller search band, the overlaps between the recovered and the true signals were once again extremely good and higher than 0.95 for all binaries.

\begin{figure}
\includegraphics[scale=0.4]{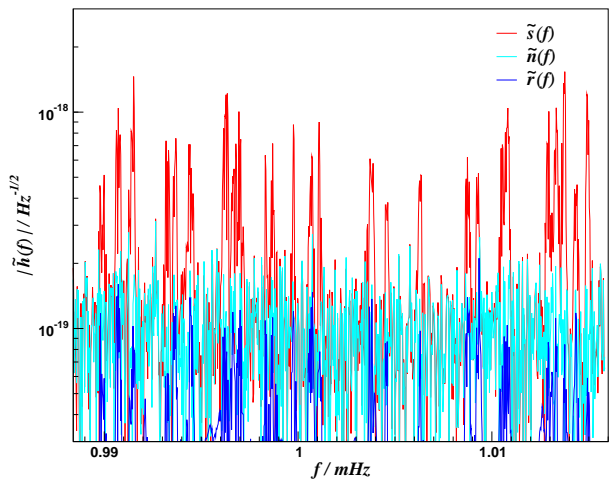}
\caption{A plot of the power spectra for the injected data set, the instrumental noise and the residual for data set 2.}
\label{fig:Set2_Residual}
\end{figure}

In Fig.~\ref{fig:Set2_Residual}, we again see that the residual power is below the noise, suggesting a minimal disturbance of the data set during source subtraction. And once again, running a sanity-test search on the source subtracted data set, we found that all simulations converged to a noise peak.  Moreover, in this case, as the
noise realization was different,  all of our simulations converged to the same frequency peak, which returned an SNR value of $\rho = 5.52$, clearly below the threshold
for detection.

%shares the signals contained in the original power and is below the noise suggesting that the substraction is good. As for the first data set, a closer insection suggest that the efficiency of the substraction depends on how well we recovered the source at the end. For binary 25, which is one of the best recovered binaries, the peak of the noise is $a.68 \times 10^{-19}$ for a residual peak of $1.20 \times 10^{-19}$. Another binary of interest is binary 2 whose recovered parameters are also close to the true values. In this case the residual is at the same level then the noise. The difference in this case  is due to the discrepancy in the values of $\phi_{0}$ and $\psi$ that are poorly recoverd. In fact, the values of the MLE and recovered values for these parameters are far away from the true values and the standard deviations from the chains are quite high $\sigma_{\phi_{0}} = 1.849$ and $\sigma_{\psi} = 0.915$.

\section{Conclusion}
In this study, we have developed a hybrid swarm-based algorithm for the detection of gravitational waves emitted by ultra compact galactic binaries, mixing together Evolutionary Algorithms such as Particle Swarm Optimization and Differential Evolution. We demonstrated the ability of this algorithm to detect a single source on a 1-mHz band using a fiducial low frequency source and the verification binary RXJ0806.3+1527 in the framework of the future eLISA mission. We then applied this search algorithm to two data sets containing 18 and 30 sources on frequency bands equal to 1 mHz (no confusion) and $10^{3} f_{m} \approx 30 \mu \text{Hz}$ 
(mild confusion) respectively. We successfully recovered all sources, and using a full Bayesian analysis, we demonstrated that the injected values were almost always within a 99\% credible interval around the median values for the recovered binaries. 

This demonstrated, that while unoptimized, the algorithm works well in the iterative search for GB sources.  We have now started
work on optimization of this version of the algorithm, and have also begun an investigation of the strong source confusion case, as well as an extension of the algorithm to the application of a simultaneous multiple-swarm search.

\appendix
\section{F-statistic}
\label{F_stat_appendix}
The F-Statistic is a method for analytically maximizing over certain parameters in the GW response.  As stated in the main body, for compact galactic binaries, we 
can split the set of waveform parameters into sets that we call intrinsic, i.e. $\{f_0,\theta,\phi\}$, and extrinsic, i.e. $\{\iota, \phi_0, A, \psi\}$.  In the LFA, we can rewrite
the detector response in Eq.~\eqref{WaweformEq} in such a way that it is expressed as a sum of constant amplitudes $a_{i}$, depending only on extrinsic parameters,
 and time varying functions $A^{i}(t)$, depending only on intrinsic parameters; i.e.,
\begin{equation}
h (t) = \sum_{i=1}^{4} a_{i} A^{i},
\end{equation}
where we define the quantities $a_{i}$ and $A_{i}$ by
\begin{eqnarray}
a_{1} &=& \dfrac{A}{2} ( (1 + \cos^{2} i ) \cos \varphi_{0} \cos 2 \psi - 2 \cos i \sin \varphi_{0} \sin 2 \psi ), \nonumber\\
a_{2} &=&- \dfrac{A}{2} ( 2 \cos i \sin \varphi_{0} \cos 2 \psi  + (1 + \cos^{2} i ) \cos \varphi_{0} \sin 2 \psi ), \nonumber\\
a_{3} &=& - \dfrac{A}{2} ( 2 \cos i \cos \varphi_{0} \sin 2 \psi  + (1 + \cos^{2} i ) \sin \varphi_{0} \cos 2 \psi ), \nonumber\\
a_{4} &=&  \dfrac{A}{2} ( (1 + \cos^{2} i ) \sin \varphi_{0} \sin 2 \psi - 2 \cos i \cos \varphi_{0} \cos 2 \psi ),\nonumber\\
\end{eqnarray}
and
\begin{eqnarray}
A^{1} &=& D^{+} (t,\theta,\phi) \cos \Phi (t, f_{0}, \theta, \phi),\nonumber \\
A^{2} &=& D^{\times} (t,\theta,\phi) \cos \Phi (t, f_{0}, \theta, \phi), \nonumber\\
A^{3} &=& D^{+} (t,\theta,\phi) \sin \Phi (t, f_{0}, \theta, \phi), \nonumber\\
A^{4} &=& D^{\times} (t,\theta,\phi) \sin \Phi (t, f_{0}, \theta, \phi). \nonumber\\
\end{eqnarray}
If we define a set of four constants $N^{i} =  \left< s | A^{i}\right>$, where $s$ is the detector output, we can solve for the $a_i$'s according to
\begin{equation}
a_{i} = M^{-1}_{i j} N^{j},
\end{equation}
where the matrix $M^{ij}$ is defined by
\beq
M^{ij} =  \left< A^{i} | A^{j} \right>.
\eeq
If we now substitute these terms into the expression for the reduced log-likelihood
\beq
\ln {\mathcal L}(\lm) = \left<s|h(\lm)\right> - \frac{1}{2}\left<h(\lm)|h(\lm)\right>,
\eeq
we can define the F-Statistic
\begin{equation}
\mathcal{F} = \log \mathcal{L} = \dfrac{1}{2} M^{-1}_{i j} N^{i} N^{j},
\label{F-stat_def}
\end{equation}
which automatically maximizes over the extrinsic parameters.

Now, given the numerical values of the amplitudes $a_{i}$, we can solve for the extrinsic parameters according to 
\begin{eqnarray}
A &=& \dfrac{A_{+} + \sqrt{A_{+}^{2} - A_{\times}^{2}}}{2}, \\
\psi &=& \dfrac{1}{2} \text{arctan} \left( \dfrac{A_{+} a_{4} - A_{\times} a_{1} }{ - ( A_{\times} a_{2} + A_{+} a_{3}) }  \right), \\
i &=& \text{arccos} \left( \dfrac{- A_{\times}}{A_{+} + \sqrt{A_{+}^{2} - A_{\times}^{2}}}     \right) , \\
\varphi_{0} &=&  \text{arctan} \left(\dfrac{ c( A_{+} a_{4} - A_{\times} a_{1}) }{ - c ( A_{+} a_{2} + A_{\times} a_{3}) }  \right),
\end{eqnarray}
where\begin{eqnarray}
A_{+} &=& \sqrt{ (a_{1} + a_{4} )^{2} + (a_{2} - a_{3})^{2} }, \nonumber\\
&+& \sqrt{ (a_{1} - a_{4} )^{2} + (a_{2} + a_{3})^{2} }, \\
A_{\times} &=& \sqrt{ (a_{1} + a_{4} )^{2} + (a_{2} - a_{3})^{2} }, \nonumber\\
&-& \sqrt{ (a_{1} - a_{4} )^{2} + (a_{2} + a_{3})^{2} } ,\\
c &=& \dfrac{ \sin(2\psi)}{\mid \sin(2\psi) \mid}.
\end{eqnarray}

As with the waveform generation, one can optimize the computation time by evaluating the functions $A^{i}$ directly in the Fourier domain.  Once again the F-statistic computed directly in the Fourier domain is faster than that in the time domain by factors of up to 30. 

\section{Tables of recovered values}
\label{Table_appendix}
Here we present the injected and recovered values from the search pipeline for both data sets.  For each each binary, the 
top row represents the injected values, and the bottom row the recovered values.

\begin{table*}\begin{center}
\scriptsize
\begin{tabular}{|c|c|c|c|c|c|c|c|c|}
\hline
Binary & $\iota$ (rad) & $\phi_{0}$ (rad) & A ($\times 10^{-22}$)  & $\psi$ (rad) & $f_{0}$ (mHz) & $\theta$ (rad) & $\phi$ (rad)  & $\rho$  \\
\hline
\multirow{2}{*}{1} & 1.921 & 5.401 & 8.592 & 2.272 & 1.2868381 & 1.607 & 0.861 & 28.574 \\ 
& 1.925 & 5.425 & 8.513 & 2.278 & 1.2868384 & 1.636 & 0.860 & 28.477 \\ 
\hline 
\multirow{2}{*}{2} & 2.037 & 0.407 & 6.234 & 2.130 & 1.4108256 & 2.109 & 1.005 & 27.049 \\ 
& 1.928 & 0.073 & 6.887 & 2.095 & 1.4108281 & 2.014 & 1.011 & 27.300 \\ 
\hline 
\multirow{2}{*}{3} & 2.614 & 5.399 & 7.001 & 1.997 & 1.0496413 & 0.657 & 4.379 & 25.834 \\ 
& 2.258 & 3.236 & 9.175 & 1.611 & 1.0496403 & 0.670 & 4.302 & 25.807 \\ 
\hline 
\multirow{2}{*}{4} & 0.973 & 3.079 & 3.562 & 2.159 & 1.4499807 & 1.822 & 2.603 & 21.913 \\ 
& 0.653 & 3.163 & 3.134 & 1.562 & 1.4499779 & 1.919 & 2.549 & 22.433 \\ 
\hline 
\multirow{2}{*}{5} & 0.794 & 1.146 & 7.543 & 2.447 & 0.9077304 & 1.899 & 3.754 & 18.679 \\ 
& 1.027 & 0.661 & 9.381 & 2.498 & 0.9077328 & 2.005 & 3.712 & 18.998 \\ 
\hline 
\multirow{2}{*}{6} & 0.169 & 1.721 & 4.138 & 2.258 & 1.0125854 & 2.527 & 3.814 & 17.991 \\ 
& 1.229 & 2.567 & 9.616 & 1.386 & 1.0125906 & 2.546 & 3.472 & 18.413 \\ 
\hline 
\multirow{2}{*}{7} & 0.707 & 5.536 & 7.505 & 0.378 & 0.8900178 & 2.181 & 2.625 & 17.933 \\ 
& 0.736 & 3.146 & 7.238 & 1.572 & 0.8900200 & 2.120 & 2.575 & 18.049 \\ 
\hline 
\multirow{2}{*}{8} & 1.374 & 5.872 & 6.656 & 1.307 & 1.1128036 & 2.440 & 2.393 & 14.569 \\ 
& 1.469 & 3.157 & 8.624 & 2.892 & 1.1128001 & 2.445 & 2.528 & 15.000 \\ 
\hline 
\multirow{2}{*}{9} & 2.265 & 5.416 & 4.265 & 2.420 & 1.0739092 & 1.861 & 2.984 & 14.004 \\ 
& 1.894 & 4.639 & 6.736 & 2.248 & 1.0739122 & 1.966 & 3.106 & 14.902 \\ 
\hline 
\multirow{2}{*}{10} & 1.334 & 5.966 & 6.307 & 0.062 & 1.1657974 & 0.910 & 4.236 & 11.869 \\ 
& 1.200 & 5.440 & 4.746 & 0.123 & 1.1658040 & 0.681 & 4.120 & 12.613 \\ 
\hline 
\multirow{2}{*}{11} & 1.292 & 6.011 & 8.820 & 2.480 & 0.9523456 & 1.860 & 5.740 & 12.057 \\ 
& 1.102 & 6.610 & 6.147 & 2.324 & 0.9523423 & 2.015 & 5.588 & 12.619 \\ 
\hline 
\multirow{2}{*}{12} & 1.505 & 1.234 & 9.139 & 2.217 & 0.8539045 & 1.698 & 3.036 & 11.698 \\ 
& 1.433 & 4.555 & 9.850 & 0.661 & 0.8539042 & 1.762 & 3.109 & 11.809 \\ 
\hline 
\multirow{2}{*}{13} & 1.310 & 1.508 & 2.749 & 0.964 & 1.4916823 & 1.821 & 3.124 & 11.179 \\ 
& 1.382 & 2.009 & 3.192 & 1.078 & 1.4916762 & 1.835 & 3.163 & 11.591 \\ 
\hline 
\multirow{2}{*}{14} & 1.037 & 0.901 & 2.971 & 0.590 & 1.2312381 & 0.796 & 0.794 & 11.129 \\ 
& 1.076 & 3.150 & 3.480 & 1.598 & 1.2312356 & 0.785 & 0.722 & 11.516 \\ 
\hline 
\multirow{2}{*}{15} & 1.633 & 2.941 & 7.499 & 1.073 & 1.0349592 & 2.173 & 4.157 & 10.819 \\ 
& 1.639 & 6.147 & 7.950 & 2.585 & 1.0349597 & 2.198 & 4.107 & 11.042 \\ 
\hline 
\multirow{2}{*}{16} & 1.234 & 1.502 & 3.227 & 2.230 & 1.2654466 & 1.224 & 0.927 & 11.207 \\ 
& 1.290 & 1.607 & 4.010 & 2.000 & 1.2654471 & 1.264 & 0.955 & 11.755 \\ 
\hline 
\multirow{2}{*}{17} & 1.885 & 0.241 & 6.627 & 1.720 & 0.8813608 & 2.248 & 6.083 & 10.498 \\ 
& 1.805 & 2.752 & 9.734 & 0.053 & 0.8813639 & 2.244 & 6.218 & 10.917 \\ 
\hline 
\multirow{2}{*}{18} & 1.642 & 2.812 & 4.951 & 1.350 & 1.3962156 & 2.417 & 6.095 & 10.585 \\ 
& 1.389 & 3.152 & 3.461 & 0.663 & 1.3961841 & 1.188 & 1.141 & 10.881 \\ 
\hline 
\end{tabular}
\caption{True values (upper line) and recovered median values (lower line) for data set 1.}
\label{tab:set1RecoveredValues}
\end{center}\end{table*}

\begin{table*}\begin{center}
\scriptsize
\begin{tabular}{|c|c|c|c|c|c|c|c|c|}
\hline
Binary & i (rad) & $\phi_{0}$ (rad) & A ($\times 10^{-22}$)  & $\psi$ (rad) & $f_{0}$ (mHz) & $\theta$ (rad) & $\phi$ (rad) & $\rho$  \\
\hline
\multirow{2}{*}{1} & 2.724 & 2.486 & 9.190 & 0.071 & 1.0137261 & 2.486 & 3.997 & 33.372 \\ 
& 2.497 & 3.161 & 10.258 & 1.565 & 1.0137248 & 2.467 & 3.987 & 33.398 \\ 
\hline 
\multirow{2}{*}{2} & 0.657 & 5.360 & 9.485 & 0.157 & 0.9914607 & 1.007 & 4.245 & 31.033 \\ 
& 0.634 & 3.157 & 9.512 & 1.569 & 0.9914620 & 0.970 & 4.190 & 31.117 \\ 
\hline 
\multirow{2}{*}{3} & 0.490 & 0.466 & 8.480 & 2.625 & 1.0149462 & 2.510 & 3.019 & 29.325 \\ 
& 0.759 & 3.130 & 9.926 & 1.599 & 1.0149459 & 2.490 & 3.027 & 29.347 \\ 
\hline 
\multirow{2}{*}{4} & 0.654 & 5.595 & 8.554 & 0.232 & 1.0107870 & 1.179 & 4.783 & 29.059 \\ 
& 0.656 & 3.140 & 8.878 & 1.571 & 1.0107844 & 1.234 & 4.760 & 29.134 \\ 
\hline 
\multirow{2}{*}{5} & 2.710 & 5.245 & 6.090 & 1.250 & 0.9907225 & 1.653 & 4.248 & 26.531 \\ 
& 2.499 & 3.159 & 7.889 & 1.568 & 0.9907217 & 1.688 & 4.258 & 26.581 \\ 
\hline 
\multirow{2}{*}{6} & 0.087 & 3.574 & 6.578 & 1.917 & 0.9879189 & 1.517 & 1.154 & 25.948 \\ 
& 0.761 & 3.098 & 8.463 & 1.604 & 0.9879187 & 1.502 & 1.136 & 25.968 \\ 
\hline 
\multirow{2}{*}{7} & 0.642 & 0.205 & 6.039 & 1.474 & 1.0129264 & 1.694 & 2.440 & 22.489 \\ 
& 0.685 & 3.125 & 6.448 & 1.560 & 1.0129254 & 1.804 & 2.430 & 22.519 \\ 
\hline 
\multirow{2}{*}{8} & 2.435 & 1.804 & 5.927 & 1.392 & 0.9962224 & 1.614 & 5.307 & 21.139 \\ 
& 2.363 & 3.160 & 6.690 & 1.579 & 0.9962234 & 1.575 & 5.335 & 21.217 \\ 
\hline 
\multirow{2}{*}{9} & 0.416 & 1.879 & 5.615 & 2.911 & 1.0132963 & 1.057 & 0.906 & 21.055 \\ 
& 0.692 & 3.127 & 6.224 & 1.600 & 1.0132978 & 1.079 & 0.948 & 20.871 \\ 
\hline 
\multirow{2}{*}{10} & 0.573 & 0.381 & 5.691 & 1.367 & 0.9969600 & 1.705 & 5.894 & 18.579 \\ 
& 0.993 & 0.325 & 7.371 & 1.546 & 0.9969575 & 1.684 & 5.814 & 18.698 \\ 
\hline 
\multirow{2}{*}{11} & 1.109 & 6.107 & 8.611 & 2.887 & 0.9932594 & 0.991 & 3.671 & 18.766 \\ 
& 1.128 & 6.293 & 9.225 & 2.788 & 0.9932599 & 0.988 & 3.661 & 18.778 \\ 
\hline 
\multirow{2}{*}{12} & 2.918 & 3.411 & 5.474 & 1.230 & 0.9936598 & 2.416 & 3.223 & 18.298 \\ 
& 2.291 & 3.209 & 6.835 & 1.601 & 0.9936605 & 2.372 & 3.157 & 18.344 \\ 
\hline 
\multirow{2}{*}{13} & 0.587 & 4.908 & 5.728 & 1.846 & 1.0010903 & 2.781 & 4.516 & 17.407 \\ 
& 0.866 & 3.097 & 6.929 & 1.525 & 1.0010910 & 2.839 & 4.481 & 17.517 \\ 
\hline 
\multirow{2}{*}{14} & 2.909 & 5.736 & 4.362 & 1.583 & 1.0105097 & 1.915 & 4.873 & 17.207 \\ 
& 2.414 & 3.059 & 5.489 & 1.617 & 1.0105107 & 1.989 & 4.873 & 17.292 \\ 
\hline 
\multirow{2}{*}{15} & 2.355 & 2.960 & 6.865 & 0.269 & 0.9997605 & 2.801 & 5.002 & 16.237 \\ 
& 2.090 & 3.164 & 8.168 & 1.543 & 0.9997629 & 2.842 & 5.003 & 16.388 \\ 
\hline 
\multirow{2}{*}{16} & 0.763 & 5.151 & 5.300 & 3.043 & 1.0006721 & 1.475 & 4.290 & 15.903 \\ 
& 1.110 & 4.872 & 7.285 & 3.084 & 1.0006747 & 1.566 & 4.321 & 16.135 \\ 
\hline 
\multirow{2}{*}{17} & 1.382 & 4.964 & 8.471 & 1.291 & 1.0087702 & 1.115 & 0.025 & 15.684 \\ 
& 1.334 & 4.871 & 9.959 & 1.292 & 1.0087716 & 1.062 & 0.017 & 15.791 \\ 
\hline 
\multirow{2}{*}{18} & 2.413 & 4.630 & 4.752 & 0.491 & 0.9867635 & 1.230 & 4.405 & 15.853 \\ 
& 2.157 & 4.796 & 6.555 & 0.566 & 0.9867638 & 1.168 & 4.404 & 15.931 \\ 
\hline 
\multirow{2}{*}{19} & 2.226 & 0.092 & 5.368 & 2.827 & 1.0063174 & 2.011 & 0.955 & 15.081 \\ 
& 2.384 & 3.134 & 4.746 & 1.570 & 1.0063165 & 2.058 & 0.949 & 15.104 \\ 
\hline 
\multirow{2}{*}{20} & 0.714 & 1.599 & 5.244 & 0.996 & 0.9849761 & 1.785 & 0.078 & 15.473 \\ 
& 0.749 & 3.038 & 5.301 & 1.501 & 0.9849765 & 1.681 & 0.068 & 15.604 \\ 
\hline 
\multirow{2}{*}{21} & 2.519 & 5.564 & 4.728 & 0.527 & 0.9966749 & 1.188 & 6.056 & 14.957 \\ 
& 2.179 & 3.109 & 5.907 & 1.568 & 0.9966725 & 1.215 & 6.029 & 14.838 \\ 
\hline 
\multirow{2}{*}{22} & 1.400 & 3.590 & 9.501 & 2.058 & 1.0037884 & 1.387 & 1.419 & 15.368 \\ 
& 1.354 & 0.594 & 9.750 & 0.408 & 1.0037880 & 1.242 & 1.459 & 15.643 \\ 
\hline 
\multirow{2}{*}{23} & 2.336 & 1.589 & 5.933 & 1.244 & 0.9944868 & 2.164 & 2.792 & 14.590 \\ 
& 2.314 & 3.114 & 5.099 & 1.569 & 0.9944899 & 2.131 & 2.710 & 14.644 \\ 
\hline 
\multirow{2}{*}{24} & 1.068 & 0.755 & 6.286 & 0.107 & 0.9874341 & 1.749 & 2.011 & 14.236 \\ 
& 0.822 & 3.113 & 4.832 & 1.572 & 0.9874327 & 1.726 & 2.039 & 14.274 \\ 
\hline 
\multirow{2}{*}{25} & 1.484 & 4.598 & 7.791 & 1.088 & 0.9987247 & 1.759 & 3.263 & 13.653 \\ 
& 1.505 & 1.332 & 9.130 & 2.646 & 0.9987257 & 1.660 & 3.227 & 13.795 \\ 
\hline 
\multirow{2}{*}{26} & 0.598 & 0.256 & 3.732 & 0.489 & 1.0143321 & 1.709 & 4.371 & 13.398 \\ 
& 0.859 & 3.167 & 4.538 & 1.584 & 1.0143343 & 1.768 & 4.295 & 13.587 \\ 
\hline 
\multirow{2}{*}{27} & 2.598 & 5.467 & 3.224 & 0.452 & 1.0092818 & 0.807 & 5.125 & 12.370 \\ 
& 2.172 & 3.135 & 5.185 & 1.577 & 1.0092834 & 0.678 & 5.351 & 12.774 \\ 
\hline 
\multirow{2}{*}{28} & 1.354 & 4.636 & 7.251 & 1.196 & 0.9898903 & 1.722 & 0.786 & 12.877 \\ 
& 1.236 & 5.007 & 6.642 & 1.146 & 0.9898880 & 1.816 & 0.754 & 13.043 \\ 
\hline 
\multirow{2}{*}{29} & 1.648 & 0.571 & 7.034 & 0.244 & 0.9983514 & 0.642 & 2.039 & 12.022 \\ 
& 1.578 & 3.455 & 7.034 & 1.850 & 0.9983527 & 0.631 & 2.197 & 12.055 \\ 
\hline 
\multirow{2}{*}{30} & 2.713 & 4.060 & 3.253 & 0.771 & 1.0045681 & 2.501 & 5.855 & 10.966 \\ 
& 2.222 & 3.121 & 4.273 & 1.584 & 1.0045687 & 2.517 & 5.928 & 10.968 \\ 
\hline 
\end{tabular}
\caption{True values (upper line) and recovered median values (lower line) for data set 2.}
\label{tab:set2RecoveredValues}
\end{center}\end{table*}

\bibliography{ArticlePSO_v7.bib}

\begin{thebibliography}{51}
\expandafter\ifx\csname natexlab\endcsname\relax\def\natexlab#1{#1}\fi
\expandafter\ifx\csname bibnamefont\endcsname\relax
  \def\bibnamefont#1{#1}\fi
\expandafter\ifx\csname bibfnamefont\endcsname\relax
  \def\bibfnamefont#1{#1}\fi
\expandafter\ifx\csname citenamefont\endcsname\relax
  \def\citenamefont#1{#1}\fi
\expandafter\ifx\csname url\endcsname\relax
  \def\url#1{\texttt{#1}}\fi
\expandafter\ifx\csname urlprefix\endcsname\relax\def\urlprefix{URL }\fi
\providecommand{\bibinfo}[2]{#2}
\providecommand{\eprint}[2][]{\url{#2}}

\bibitem[{\citenamefont{Nelemans et~al.}(2001)\citenamefont{Nelemans,
  Yungelson, and Portegies~Zwart}}]{nelemans_2001}
\bibinfo{author}{\bibfnamefont{G.}~\bibnamefont{Nelemans}},
  \bibinfo{author}{\bibfnamefont{L.~R.} \bibnamefont{Yungelson}},
  \bibnamefont{and} \bibinfo{author}{\bibfnamefont{S.~F.}
  \bibnamefont{Portegies~Zwart}}, \bibinfo{journal}{Astron. Astrophys.}
  \textbf{\bibinfo{volume}{375}}, \bibinfo{pages}{890} (\bibinfo{year}{2001}).

\bibitem[{\citenamefont{Ruiter et~al.}(2010)\citenamefont{Ruiter, Belczynski,
  Benacquista, Larson, and Williams}}]{ruiter_2010}
\bibinfo{author}{\bibfnamefont{A.~J.} \bibnamefont{Ruiter}},
  \bibinfo{author}{\bibfnamefont{K.}~\bibnamefont{Belczynski}},
  \bibinfo{author}{\bibfnamefont{M.}~\bibnamefont{Benacquista}},
  \bibinfo{author}{\bibfnamefont{S.~L.} \bibnamefont{Larson}},
  \bibnamefont{and} \bibinfo{author}{\bibfnamefont{G.}~\bibnamefont{Williams}},
  \bibinfo{journal}{Astrophys.J.} \textbf{\bibinfo{volume}{717}},
  \bibinfo{pages}{1006} (\bibinfo{year}{2010}).

\bibitem[{\citenamefont{Nelemans}(2013)}]{Nelemans:2013yg}
\bibinfo{author}{\bibfnamefont{G.}~\bibnamefont{Nelemans}},
  \bibinfo{journal}{ASP Conf. Ser.} \textbf{\bibinfo{volume}{467}},
  \bibinfo{pages}{27} (\bibinfo{year}{2013}).

\bibitem[{\citenamefont{Nelemans et~al.}(2010)\citenamefont{Nelemans, Wood,
  Groot, Anderson, Belczynski, Benacquista, Charles, Cumming, Deloye, Jonker
  et~al.}}]{nelemans_astrophysics_2009}
\bibinfo{author}{\bibfnamefont{G.}~\bibnamefont{Nelemans}},
  \bibinfo{author}{\bibfnamefont{M.}~\bibnamefont{Wood}},
  \bibinfo{author}{\bibfnamefont{P.}~\bibnamefont{Groot}},
  \bibinfo{author}{\bibfnamefont{S.}~\bibnamefont{Anderson}},
  \bibinfo{author}{\bibfnamefont{K.}~\bibnamefont{Belczynski}},
  \bibinfo{author}{\bibfnamefont{M.}~\bibnamefont{Benacquista}},
  \bibinfo{author}{\bibfnamefont{P.}~\bibnamefont{Charles}},
  \bibinfo{author}{\bibfnamefont{A.}~\bibnamefont{Cumming}},
  \bibinfo{author}{\bibfnamefont{C.}~\bibnamefont{Deloye}},
  \bibinfo{author}{\bibfnamefont{P.}~\bibnamefont{Jonker}},
  \bibnamefont{et~al.}, \bibinfo{journal}{White paper for the National
  Academics Astro 2010}  (\bibinfo{year}{2010}).

\bibitem[{\citenamefont{Kilic et~al.}(2013)\citenamefont{Kilic, Brown, and
  Hermes}}]{kilic_ultra-compact_2013}
\bibinfo{author}{\bibfnamefont{M.}~\bibnamefont{Kilic}},
  \bibinfo{author}{\bibfnamefont{W.~R.} \bibnamefont{Brown}}, \bibnamefont{and}
  \bibinfo{author}{\bibfnamefont{J.~J.} \bibnamefont{Hermes}}, in
  \emph{\bibinfo{booktitle}{9th {LISA} {Symposium}, {Astronomical} {Society} of
  the {Pacific} {Conference} {Series}}} (\bibinfo{year}{2013}), vol.
  \bibinfo{volume}{467}, p.~\bibinfo{pages}{47}.

\bibitem[{\citenamefont{Cornish and Crowder}(2005)}]{cornish_lisa_2005}
\bibinfo{author}{\bibfnamefont{N.~J.} \bibnamefont{Cornish}} \bibnamefont{and}
  \bibinfo{author}{\bibfnamefont{J.}~\bibnamefont{Crowder}},
  \bibinfo{journal}{Phys.Rev.D} \textbf{\bibinfo{volume}{72}},
  \bibinfo{pages}{043005} (\bibinfo{year}{2005}).

\bibitem[{\citenamefont{Amaro-Seoane et~al.}(2012)\citenamefont{Amaro-Seoane,
  Aoudia, Babak, Binétruy, Berti, Bohe, Caprini, Colpi, Cornish, Danzmann
  et~al.}}]{amaro-seoane_elisa:_2012}
\bibinfo{author}{\bibfnamefont{P.}~\bibnamefont{Amaro-Seoane}},
  \bibinfo{author}{\bibfnamefont{S.}~\bibnamefont{Aoudia}},
  \bibinfo{author}{\bibfnamefont{S.}~\bibnamefont{Babak}},
  \bibinfo{author}{\bibfnamefont{P.}~\bibnamefont{Binétruy}},
  \bibinfo{author}{\bibfnamefont{E.}~\bibnamefont{Berti}},
  \bibinfo{author}{\bibfnamefont{A.}~\bibnamefont{Bohe}},
  \bibinfo{author}{\bibfnamefont{C.}~\bibnamefont{Caprini}},
  \bibinfo{author}{\bibfnamefont{M.}~\bibnamefont{Colpi}},
  \bibinfo{author}{\bibfnamefont{N.~J.} \bibnamefont{Cornish}},
  \bibinfo{author}{\bibfnamefont{K.}~\bibnamefont{Danzmann}},
  \bibnamefont{et~al.}, \bibinfo{journal}{Classical Quantum Gravity}
  \textbf{\bibinfo{volume}{29}}, \bibinfo{pages}{124016}
  (\bibinfo{year}{2012}).

\bibitem[{\citenamefont{Cornish and Larson}(2003)}]{cornish_lisa_2003}
\bibinfo{author}{\bibfnamefont{N.~J.} \bibnamefont{Cornish}} \bibnamefont{and}
  \bibinfo{author}{\bibfnamefont{S.~L.} \bibnamefont{Larson}},
  \bibinfo{journal}{Phys.Rev.D} \textbf{\bibinfo{volume}{67}},
  \bibinfo{pages}{103001} (\bibinfo{year}{2003}).

\bibitem[{\citenamefont{Blaut et~al.}(2010)\citenamefont{Blaut, Babak, and
  Krolak}}]{Blaut:2009si}
\bibinfo{author}{\bibfnamefont{A.}~\bibnamefont{Blaut}},
  \bibinfo{author}{\bibfnamefont{S.}~\bibnamefont{Babak}}, \bibnamefont{and}
  \bibinfo{author}{\bibfnamefont{A.}~\bibnamefont{Krolak}},
  \bibinfo{journal}{Phys.Rev.D} \textbf{\bibinfo{volume}{81}},
  \bibinfo{pages}{063008} (\bibinfo{year}{2010}).

\bibitem[{\citenamefont{Whelan et~al.}(2010)\citenamefont{Whelan, Prix, and
  Khurana}}]{whelan_searching_2010}
\bibinfo{author}{\bibfnamefont{J.~T.} \bibnamefont{Whelan}},
  \bibinfo{author}{\bibfnamefont{R.}~\bibnamefont{Prix}}, \bibnamefont{and}
  \bibinfo{author}{\bibfnamefont{D.}~\bibnamefont{Khurana}},
  \bibinfo{journal}{Classical Quantum Gravity} \textbf{\bibinfo{volume}{27}},
  \bibinfo{pages}{055010} (\bibinfo{year}{2010}).

\bibitem[{\citenamefont{Cornish and Porter}(2005)}]{cornish_detecting_2005}
\bibinfo{author}{\bibfnamefont{N.~J.} \bibnamefont{Cornish}} \bibnamefont{and}
  \bibinfo{author}{\bibfnamefont{E.~K.} \bibnamefont{Porter}},
  \bibinfo{journal}{Classical Quantum Gravity} \textbf{\bibinfo{volume}{22}},
  \bibinfo{pages}{S927} (\bibinfo{year}{2005}).

\bibitem[{\citenamefont{Crowder et~al.}(2006)\citenamefont{Crowder, Cornish,
  and Reddinger}}]{crowder_lisa_2006}
\bibinfo{author}{\bibfnamefont{J.}~\bibnamefont{Crowder}},
  \bibinfo{author}{\bibfnamefont{N.~J.} \bibnamefont{Cornish}},
  \bibnamefont{and}
  \bibinfo{author}{\bibfnamefont{L.}~\bibnamefont{Reddinger}},
  \bibinfo{journal}{Phys.Rev.D} \textbf{\bibinfo{volume}{73}},
  \bibinfo{pages}{063011} (\bibinfo{year}{2006}).

\bibitem[{\citenamefont{Mohanty and Nayak}(2006)}]{mohanty_tomographic_2005}
\bibinfo{author}{\bibfnamefont{S.~D.} \bibnamefont{Mohanty}} \bibnamefont{and}
  \bibinfo{author}{\bibfnamefont{R.~K.} \bibnamefont{Nayak}},
  \bibinfo{journal}{Phys.Rev.D} \textbf{\bibinfo{volume}{73}},
  \bibinfo{pages}{083006} (\bibinfo{year}{2006}).

\bibitem[{\citenamefont{Trias et~al.}(2009)\citenamefont{Trias, Vecchio, and
  Veitch}}]{trias_studying_2009}
\bibinfo{author}{\bibfnamefont{M.}~\bibnamefont{Trias}},
  \bibinfo{author}{\bibfnamefont{A.}~\bibnamefont{Vecchio}}, \bibnamefont{and}
  \bibinfo{author}{\bibfnamefont{J.}~\bibnamefont{Veitch}},
  \bibinfo{journal}{Classical Quantum Gravity} \textbf{\bibinfo{volume}{26}},
  \bibinfo{pages}{204024} (\bibinfo{year}{2009}).

\bibitem[{\citenamefont{Crowder and Cornish}(2007)}]{crowder_solution_2006}
\bibinfo{author}{\bibfnamefont{J.}~\bibnamefont{Crowder}} \bibnamefont{and}
  \bibinfo{author}{\bibfnamefont{N.}~\bibnamefont{Cornish}},
  \bibinfo{journal}{Phys.Rev.D} \textbf{\bibinfo{volume}{75}},
  \bibinfo{pages}{043008} (\bibinfo{year}{2007}).

\bibitem[{\citenamefont{Cornish and
  Porter}(2007{\natexlab{a}})}]{cornish_porter_2007_2}
\bibinfo{author}{\bibfnamefont{N.~J.} \bibnamefont{Cornish}} \bibnamefont{and}
  \bibinfo{author}{\bibfnamefont{E.~K.} \bibnamefont{Porter}},
  \bibinfo{journal}{Classical Quantum Gravity} \textbf{\bibinfo{volume}{24}},
  \bibinfo{pages}{5729} (\bibinfo{year}{2007}{\natexlab{a}}).

\bibitem[{\citenamefont{Cornish and
  Porter}(2007{\natexlab{b}})}]{cornish_porter_2007_3}
\bibinfo{author}{\bibfnamefont{N.~J.} \bibnamefont{Cornish}} \bibnamefont{and}
  \bibinfo{author}{\bibfnamefont{E.~K.} \bibnamefont{Porter}},
  \bibinfo{journal}{Classical Quantum Gravity} \textbf{\bibinfo{volume}{24}},
  \bibinfo{pages}{S501} (\bibinfo{year}{2007}{\natexlab{b}}).

\bibitem[{\citenamefont{Vecchio and Wickham}(2004)}]{vecchio_2004}
\bibinfo{author}{\bibfnamefont{A.}~\bibnamefont{Vecchio}} \bibnamefont{and}
  \bibinfo{author}{\bibfnamefont{E.~D.} \bibnamefont{Wickham}},
  \bibinfo{journal}{Phys.Rev.D} \textbf{\bibinfo{volume}{70}},
  \bibinfo{pages}{082002} (\bibinfo{year}{2004}).

\bibitem[{\citenamefont{Stroeer et~al.}(2006)\citenamefont{Stroeer, Gair, and
  Vecchio}}]{vecchio_2006}
\bibinfo{author}{\bibfnamefont{A.}~\bibnamefont{Stroeer}},
  \bibinfo{author}{\bibfnamefont{J.}~\bibnamefont{Gair}}, \bibnamefont{and}
  \bibinfo{author}{\bibfnamefont{A.}~\bibnamefont{Vecchio}},
  \bibinfo{journal}{AIP Conf.Proc.} \textbf{\bibinfo{volume}{873}},
  \bibinfo{pages}{444} (\bibinfo{year}{2006}).

\bibitem[{\citenamefont{Veitch and Vecchio}(2008)}]{veitch_2008}
\bibinfo{author}{\bibfnamefont{J.}~\bibnamefont{Veitch}} \bibnamefont{and}
  \bibinfo{author}{\bibfnamefont{A.}~\bibnamefont{Vecchio}},
  \bibinfo{journal}{Phys.Rev.D} \textbf{\bibinfo{volume}{78}},
  \bibinfo{pages}{022001} (\bibinfo{year}{2008}).

\bibitem[{\citenamefont{Cornish and Littenberg}(2007)}]{cornish_tests_2007}
\bibinfo{author}{\bibfnamefont{N.~J.} \bibnamefont{Cornish}} \bibnamefont{and}
  \bibinfo{author}{\bibfnamefont{T.~B.} \bibnamefont{Littenberg}},
  \bibinfo{journal}{Phys.Rev.D} \textbf{\bibinfo{volume}{76}},
  \bibinfo{pages}{083006} (\bibinfo{year}{2007}).

\bibitem[{\citenamefont{Littenberg and Cornish}(2009)}]{littenberg_2009}
\bibinfo{author}{\bibfnamefont{T.~B.} \bibnamefont{Littenberg}}
  \bibnamefont{and} \bibinfo{author}{\bibfnamefont{N.~J.}
  \bibnamefont{Cornish}}, \bibinfo{journal}{Phys.Rev.D}
  \textbf{\bibinfo{volume}{80}}, \bibinfo{pages}{063007}
  (\bibinfo{year}{2009}).

\bibitem[{\citenamefont{Kennedy and Eberhart}(1995)}]{kennedy_1995}
\bibinfo{author}{\bibfnamefont{J.}~\bibnamefont{Kennedy}} \bibnamefont{and}
  \bibinfo{author}{\bibfnamefont{R.}~\bibnamefont{Eberhart}},
  \bibinfo{journal}{IEEE International Conference on Neural Networks
  Proceedings} \textbf{\bibinfo{volume}{4}}, \bibinfo{pages}{1942}
  (\bibinfo{year}{1995}).

\bibitem[{\citenamefont{Shi and Eberhart}(1998)}]{shi_1998}
\bibinfo{author}{\bibfnamefont{Y.}~\bibnamefont{Shi}} \bibnamefont{and}
  \bibinfo{author}{\bibfnamefont{R.}~\bibnamefont{Eberhart}},
  \bibinfo{journal}{The 1998 IEEE International Conference on Evolutionary
  Computation Proceedings, IEEE Computer Society, Washington DC, USA.}
  p.~\bibinfo{pages}{69} (\bibinfo{year}{1998}).

\bibitem[{\citenamefont{Storn and Price}(1997)}]{storn_1997}
\bibinfo{author}{\bibfnamefont{R.}~\bibnamefont{Storn}} \bibnamefont{and}
  \bibinfo{author}{\bibfnamefont{K.}~\bibnamefont{Price}}, \bibinfo{journal}{J.
  Global Opt.} \textbf{\bibinfo{volume}{11}}, \bibinfo{pages}{341}
  (\bibinfo{year}{1997}).

\bibitem[{\citenamefont{{Metropolis} et~al.}(1953)\citenamefont{{Metropolis},
  {Rosenbluth}, {Rosenbluth}, {Teller}, and {Teller}}}]{metropolis_1953}
\bibinfo{author}{\bibfnamefont{N.}~\bibnamefont{{Metropolis}}},
  \bibinfo{author}{\bibfnamefont{A.~W.} \bibnamefont{{Rosenbluth}}},
  \bibinfo{author}{\bibfnamefont{M.~N.} \bibnamefont{{Rosenbluth}}},
  \bibinfo{author}{\bibfnamefont{A.~H.} \bibnamefont{{Teller}}},
  \bibnamefont{and} \bibinfo{author}{\bibfnamefont{E.}~\bibnamefont{{Teller}}},
  \bibinfo{journal}{J. Chem. Phys.} \textbf{\bibinfo{volume}{21}},
  \bibinfo{pages}{1087} (\bibinfo{year}{1953}).

\bibitem[{\citenamefont{Hastings}(1970)}]{hastings_1970}
\bibinfo{author}{\bibfnamefont{W.~K.} \bibnamefont{Hastings}},
  \bibinfo{journal}{Biometrika} \textbf{\bibinfo{volume}{57}},
  \bibinfo{pages}{97} (\bibinfo{year}{1970}).

\bibitem[{\citenamefont{Omran et~al.}(2009)\citenamefont{Omran, Engelbrecht,
  and Salman}}]{Omran_2008}
\bibinfo{author}{\bibfnamefont{M.~G.~H.} \bibnamefont{Omran}},
  \bibinfo{author}{\bibfnamefont{A.~P.} \bibnamefont{Engelbrecht}},
  \bibnamefont{and} \bibinfo{author}{\bibfnamefont{A.}~\bibnamefont{Salman}},
  \bibinfo{journal}{Eur. J. Oper. Res.} \textbf{\bibinfo{volume}{196}},
  \bibinfo{pages}{128} (\bibinfo{year}{2009}).

\bibitem[{\citenamefont{Prasad and Souradeep}(2012)}]{prasad_cosmological_2012}
\bibinfo{author}{\bibfnamefont{J.}~\bibnamefont{Prasad}} \bibnamefont{and}
  \bibinfo{author}{\bibfnamefont{T.}~\bibnamefont{Souradeep}},
  \bibinfo{journal}{Phys.Rev.D} \textbf{\bibinfo{volume}{85}},
  \bibinfo{pages}{123008} (\bibinfo{year}{2012}), \bibinfo{note}{[Erratum:
  Phys. Rev.D90,no.10,109903(2014)]}.

\bibitem[{\citenamefont{Hulse and Taylor}(1975)}]{hulse_1975}
\bibinfo{author}{\bibfnamefont{R.~A.} \bibnamefont{Hulse}} \bibnamefont{and}
  \bibinfo{author}{\bibfnamefont{J.~H.} \bibnamefont{Taylor}},
  \bibinfo{journal}{Astrophys.J.} \textbf{\bibinfo{volume}{195}},
  \bibinfo{pages}{L51} (\bibinfo{year}{1975}).

\bibitem[{\citenamefont{Aasi et~al.}(2015)}]{TheLIGOScientific:2014jea}
\bibinfo{author}{\bibfnamefont{J.}~\bibnamefont{Aasi}} \bibnamefont{et~al.}
  (\bibinfo{collaboration}{LIGO Scientific Collaboration}),
  \bibinfo{journal}{Classical Quantum Gravity} \textbf{\bibinfo{volume}{32}},
  \bibinfo{pages}{115012} (\bibinfo{year}{2015}), \eprint{1411.4547}.

\bibitem[{\citenamefont{Acernese et~al.}(2015)}]{TheVirgo:2014hva}
\bibinfo{author}{\bibfnamefont{F.}~\bibnamefont{Acernese}} \bibnamefont{et~al.}
  (\bibinfo{collaboration}{VIRGO}), \bibinfo{journal}{Classical Quantum
  Gravity} \textbf{\bibinfo{volume}{32}}, \bibinfo{pages}{024001}
  (\bibinfo{year}{2015}).

\bibitem[{pta()}]{pta}
\urlprefix\url{http://www.ipta4gw.org/}.

\bibitem[{\citenamefont{Nissanke et~al.}(2012)\citenamefont{Nissanke,
  Vallisneri, Nelemans, and Prince}}]{nissanke_gravitational-wave_2012}
\bibinfo{author}{\bibfnamefont{S.}~\bibnamefont{Nissanke}},
  \bibinfo{author}{\bibfnamefont{M.}~\bibnamefont{Vallisneri}},
  \bibinfo{author}{\bibfnamefont{G.}~\bibnamefont{Nelemans}}, \bibnamefont{and}
  \bibinfo{author}{\bibfnamefont{T.~A.} \bibnamefont{Prince}},
  \bibinfo{journal}{Astrophys.J.} \textbf{\bibinfo{volume}{758}},
  \bibinfo{pages}{131} (\bibinfo{year}{2012}).

\bibitem[{\citenamefont{Porter and Cornish}(2015)}]{porter_fisher_2015}
\bibinfo{author}{\bibfnamefont{E.~K.} \bibnamefont{Porter}} \bibnamefont{and}
  \bibinfo{author}{\bibfnamefont{N.~J.} \bibnamefont{Cornish}},
  \bibinfo{journal}{Phys.Rev.D} \textbf{\bibinfo{volume}{91}},
  \bibinfo{pages}{104001} (\bibinfo{year}{2015}).

\bibitem[{\citenamefont{Cutler}(1998)}]{cutler_1998}
\bibinfo{author}{\bibfnamefont{C.}~\bibnamefont{Cutler}},
  \bibinfo{journal}{Phys.Rev.D} \textbf{\bibinfo{volume}{57}},
  \bibinfo{pages}{7089} (\bibinfo{year}{1998}).

\bibitem[{\citenamefont{Cornish and Rubbo}(2003)}]{cornish_rubbo2003}
\bibinfo{author}{\bibfnamefont{N.~J.} \bibnamefont{Cornish}} \bibnamefont{and}
  \bibinfo{author}{\bibfnamefont{L.~J.} \bibnamefont{Rubbo}},
  \bibinfo{journal}{Phys.Rev.D} \textbf{\bibinfo{volume}{67}},
  \bibinfo{pages}{022001} (\bibinfo{year}{2003}), \bibinfo{note}{[Erratum:
  Phys. Rev.D67,029905(2003)]}.

\bibitem[{\citenamefont{Krolak et~al.}(2004)\citenamefont{Krolak, Tinto, and
  Vallisneri}}]{krolac_2004}
\bibinfo{author}{\bibfnamefont{A.}~\bibnamefont{Krolak}},
  \bibinfo{author}{\bibfnamefont{M.}~\bibnamefont{Tinto}}, \bibnamefont{and}
  \bibinfo{author}{\bibfnamefont{M.}~\bibnamefont{Vallisneri}},
  \bibinfo{journal}{Phys.Rev.D} \textbf{\bibinfo{volume}{70}},
  \bibinfo{pages}{022003} (\bibinfo{year}{2004}), \bibinfo{note}{[Erratum:
  Phys. Rev.D76,069901(2007)]}.

\bibitem[{\citenamefont{Timpano et~al.}(2006)\citenamefont{Timpano, Rubbo, and
  Cornish}}]{timpano_2006}
\bibinfo{author}{\bibfnamefont{S.~E.} \bibnamefont{Timpano}},
  \bibinfo{author}{\bibfnamefont{L.~J.} \bibnamefont{Rubbo}}, \bibnamefont{and}
  \bibinfo{author}{\bibfnamefont{N.~J.} \bibnamefont{Cornish}},
  \bibinfo{journal}{Phys.Rev.D} \textbf{\bibinfo{volume}{73}},
  \bibinfo{pages}{122001} (\bibinfo{year}{2006}).

\bibitem[{\citenamefont{Petiteau et~al.}(2010)\citenamefont{Petiteau, Shang,
  Babak, and Feroz}}]{Petiteau:2010zu}
\bibinfo{author}{\bibfnamefont{A.}~\bibnamefont{Petiteau}},
  \bibinfo{author}{\bibfnamefont{Y.}~\bibnamefont{Shang}},
  \bibinfo{author}{\bibfnamefont{S.}~\bibnamefont{Babak}}, \bibnamefont{and}
  \bibinfo{author}{\bibfnamefont{F.}~\bibnamefont{Feroz}},
  \bibinfo{journal}{Phys.Rev.D} \textbf{\bibinfo{volume}{81}},
  \bibinfo{pages}{104016} (\bibinfo{year}{2010}).

\bibitem[{\citenamefont{Gair and Porter}(2009)}]{gair_cosmic_2009}
\bibinfo{author}{\bibfnamefont{J.~R.} \bibnamefont{Gair}} \bibnamefont{and}
  \bibinfo{author}{\bibfnamefont{E.~K.} \bibnamefont{Porter}},
  \bibinfo{journal}{Classical Quantum Gravity} \textbf{\bibinfo{volume}{26}},
  \bibinfo{pages}{225004} (\bibinfo{year}{2009}).

\bibitem[{\citenamefont{ter Braak}(2006)}]{terbraak2006}
\bibinfo{author}{\bibfnamefont{C.}~\bibnamefont{ter Braak}},
  \bibinfo{journal}{Stat. Comp.} \textbf{\bibinfo{volume}{16}},
  \bibinfo{pages}{239} (\bibinfo{year}{2006}).

\bibitem[{\citenamefont{ter Braak and Vrugt}(2008)}]{terbraak2008}
\bibinfo{author}{\bibfnamefont{C.}~\bibnamefont{ter Braak}} \bibnamefont{and}
  \bibinfo{author}{\bibfnamefont{J.}~\bibnamefont{Vrugt}},
  \bibinfo{journal}{Stat. Comp.} \textbf{\bibinfo{volume}{18}},
  \bibinfo{pages}{435} (\bibinfo{year}{2008}).

\bibitem[{\citenamefont{Taylor et~al.}(2012)\citenamefont{Taylor, Gair, and
  Lentati}}]{taylor_2012}
\bibinfo{author}{\bibfnamefont{S.~R.} \bibnamefont{Taylor}},
  \bibinfo{author}{\bibfnamefont{J.~R.} \bibnamefont{Gair}}, \bibnamefont{and}
  \bibinfo{author}{\bibfnamefont{L.}~\bibnamefont{Lentati}},
  \bibinfo{journal}{arXiv:1210.3489}  (\bibinfo{year}{2012}).

\bibitem[{\citenamefont{Wang et~al.}(2014)\citenamefont{Wang, Mohanty, and
  Jenet}}]{wang_2014}
\bibinfo{author}{\bibfnamefont{Y.}~\bibnamefont{Wang}},
  \bibinfo{author}{\bibfnamefont{S.~D.} \bibnamefont{Mohanty}},
  \bibnamefont{and} \bibinfo{author}{\bibfnamefont{F.~A.} \bibnamefont{Jenet}},
  \bibinfo{journal}{Astrophys. J.} \textbf{\bibinfo{volume}{795}},
  \bibinfo{pages}{96} (\bibinfo{year}{2014}).

\bibitem[{\citenamefont{Wang and Mohanty}(2010)}]{wang_2010}
\bibinfo{author}{\bibfnamefont{Y.}~\bibnamefont{Wang}} \bibnamefont{and}
  \bibinfo{author}{\bibfnamefont{S.~D.} \bibnamefont{Mohanty}},
  \bibinfo{journal}{Phys.Rev.D} \textbf{\bibinfo{volume}{81}},
  \bibinfo{pages}{063002} (\bibinfo{year}{2010}).

\bibitem[{sta()}]{standard_PSO}
\urlprefix\url{http://www.particleswarm.info/Standard_PSO_2006.c}.

\bibitem[{\citenamefont{Huwyler et~al.}(2015)\citenamefont{Huwyler, Porter, and
  Jetzer}}]{Huwyler:2014vva}
\bibinfo{author}{\bibfnamefont{C.}~\bibnamefont{Huwyler}},
  \bibinfo{author}{\bibfnamefont{E.~K.} \bibnamefont{Porter}},
  \bibnamefont{and} \bibinfo{author}{\bibfnamefont{P.}~\bibnamefont{Jetzer}},
  \bibinfo{journal}{Phys.Rev.D} \textbf{\bibinfo{volume}{91}},
  \bibinfo{pages}{024037} (\bibinfo{year}{2015}).

\bibitem[{\citenamefont{Cornish and
  Porter}(2007{\natexlab{c}})}]{cornish_porter_2007_1}
\bibinfo{author}{\bibfnamefont{N.~J.} \bibnamefont{Cornish}} \bibnamefont{and}
  \bibinfo{author}{\bibfnamefont{E.~K.} \bibnamefont{Porter}},
  \bibinfo{journal}{Phys.Rev.D} \textbf{\bibinfo{volume}{75}},
  \bibinfo{pages}{021301} (\bibinfo{year}{2007}{\natexlab{c}}).

\bibitem[{\citenamefont{Roelofs et~al.}(2010)\citenamefont{Roelofs, Rau, Marsh,
  Steeghs, Groot, and Nelemans}}]{roelofs_2010}
\bibinfo{author}{\bibfnamefont{G.~H.~A.} \bibnamefont{Roelofs}},
  \bibinfo{author}{\bibfnamefont{A.}~\bibnamefont{Rau}},
  \bibinfo{author}{\bibfnamefont{T.~R.} \bibnamefont{Marsh}},
  \bibinfo{author}{\bibfnamefont{D.}~\bibnamefont{Steeghs}},
  \bibinfo{author}{\bibfnamefont{P.~J.} \bibnamefont{Groot}}, \bibnamefont{and}
  \bibinfo{author}{\bibfnamefont{G.}~\bibnamefont{Nelemans}},
  \bibinfo{journal}{Astrophys. J.} \textbf{\bibinfo{volume}{711}},
  \bibinfo{pages}{L138} (\bibinfo{year}{2010}).

\bibitem[{\citenamefont{Porter}(2015)}]{porter_alone_2015}
\bibinfo{author}{\bibfnamefont{E.~K.} \bibnamefont{Porter}},
  \bibinfo{journal}{Phys.Rev.D} \textbf{\bibinfo{volume}{92}},
  \bibinfo{pages}{064001} (\bibinfo{year}{2015}).

\end{thebibliography}

\end{document}